\documentclass[11pt, a4paper]{article}
\usepackage{jheppub}
 \pdfoutput=1
\usepackage{bm, mathrsfs, physics, slashed, tensor, cancel}
\usepackage{tikz, caption, subcaption, float}
\newcommand{\be}{\begin{equation}}
\newcommand{\ee}{\end{equation}} 
\newcommand{\bea}{\begin{eqnarray}}
\newcommand{\eea}{\end{eqnarray}}
\title{Holographic Lieb lattice and gapping its Dirac band}
\author{Young-Kwon Han,}
\author{Jeong-Won Seo,}
\author{Taewon Yuk,}
\author{and Sang-Jin Sin}
\affiliation{Department of Physics, Hanyang University,\\Seoul 04763, South Korea}
\emailAdd{youngkwonhan346@gmail.com}
\emailAdd{1113dino@naver.com}
\emailAdd{tae1yuk@gmail.com}
\emailAdd{sangjin.sin@gmail.com}
\abstract
{
  We first point out that the Laia-Tong model realizes the Lieb lattice in the holographic setup.
  It generates a flat band of sharp particle spectrum together with a Dirac band of unparticle spectrum. We    provided an understanding why the Laia-Tong model's boundary condition generate a flat band and compared it with the mechanism of  'compact localized orbits" in the lattice models to provide a physical reason why Lieb and Laia-Tong model should be identified based on the similarity in the flat band generation mechanism. 
  We then construct a model which opens a gap to the Dirac band so that one can realize a well-separated flat band.  We then study the phase transition between the gapped and gapless phases analytically.
  We also made methodological progress to find a few other possible quantizations and we express the Green functions in any quantization in terms of that in the standard quantization.
 Finally we carried out the problem of back reaction to show that the qualitative feature remains the same.   
}

\begin{document}

\maketitle
\flushbottom

\section{Introduction}
Recently, the flat band (FB) is attracting much interest after most strong correlation phenomena were observed in the magic-angle twisted bilayer graphene (MATBG)
\cite{Cao:2018:10.1038_nature26154, Cao:2018:10.1038_nature26160},
the first material realizing the flat band.
It forms a huge degeneracy of localized eigenstates so that even a small interaction or impurities can cause a system instability toward various strong correlation phenomena including ferromagnetism
\cite{Mielke:1991:10.1088_0305-4470_24_14_018, Mielke:1992:10.1088_0305-4470_25_16_011, Tasaki:1992:10.1103_PhysRevLett.69.1608, Lieb:1989:10.1103_PhysRevLett.62.1201, Costa:2016:10.1103_PhysRevB.94.155107, Tamura:2002:10.1103_PhysRevB.65.085324},
superconductivity
\cite{Imada:2000:10.1103_PhysRevLett.84.143, Kopnin:2011:10.1103_PhysRevB.83.220503, Julku:2016:10.1103_PhysRevLett.117.045303},
Mott insulators
\cite{Dauphin:2016:10.1103_PhysRevA.93.043611},
and fractional quantum Hall effects (FQHE)
\cite{Katsura:2010:10.1209_0295-5075_91_57007, Green:2010:10.1103_PhysRevB.82.075104, Tang:2011:10.1103_PhysRevLett.106.236802, Sun:2011:10.1103_PhysRevLett.106.236803, Neupert:2011:10.1103_PhysRevLett.106.236804, Wang:2011:10.1103_PhysRevLett.107.146803, Sheng:2011:10.1038_ncomms1380}. 
In fact, it was more than 10 years ago when it was predicted that FQHE would be realized in the flat band without the external magnetic field.
Extensive works have been done thereafter, to create artificial lattices that can generate the flat band by the geometric frustration
\cite{Weeks:2010:10.1103_PhysRevB.82.085310, Guo:2009:10.1103_PhysRevB.80.113102, Goldman:2011:10.1103_PhysRevA.83.063601, Beugeling:2012:10.1103_PhysRevB.86.195129, Tadjine:2016:10.1103_PhysRevB.94.075441, Li:2016:10.1039_C6NR03223K}.
The experimental realizations of such lattices were successful only in the optical lattices
\cite{Wu:2007:10.1103_PhysRevLett.99.070401, Apaja:2010:10.1103_PhysRevA.82.041402, Shen:2010:10.1103_PhysRevB.81.041410, Guzman-Silva:2014:10.1088_1367-2630_16_6_063061, Mukherjee:2015:10.1103_PhysRevLett.114.245504, Vicencio:2015:10.1103_PhysRevLett.114.245503, Taie:2015:10.1126_sciadv.1500854, Xia:2016:10.1364_OL.41.001435, Diebel:2016:10.1103_PhysRevLett.116.183902}
until the invention of the MATBG.
The prediction was finally fulfilled after a decade when the fractional Chern insulator was observed in the MATBG.

\begin{figure}[t]
  \centering
  \begin{subfigure}[h]{0.33 \textwidth}
    \centering
    \includegraphics[width = 1.9 in]{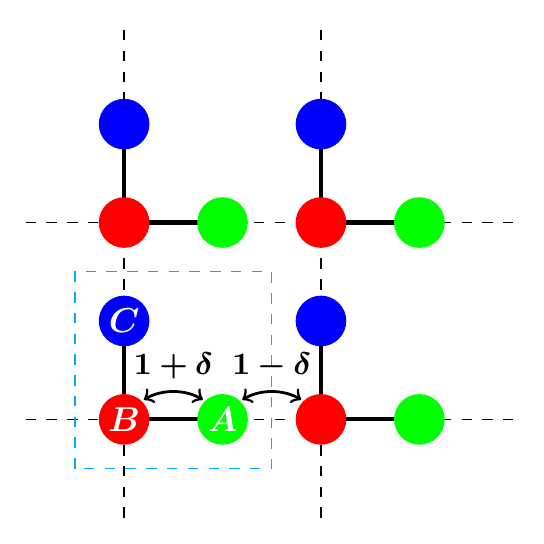}
    \caption{Lieb lattice}
    \label{subfigure:lieb_lattice_geometric_structure}
  \end{subfigure}
  \begin{subfigure}[h]{0.3 \textwidth}
    \centering
    \includegraphics[width = 1.5 in]{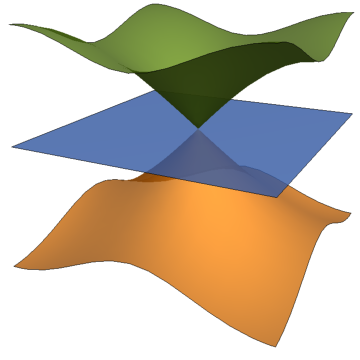}
    \caption{Gapless ($\delta = 0$)}
    \label{subfigure:lieb_lattice_gapless_energy_dispersion}
  \end{subfigure}
  \begin{subfigure}[h]{0.3 \textwidth}
    \centering
    \includegraphics[width = 1.5 in]{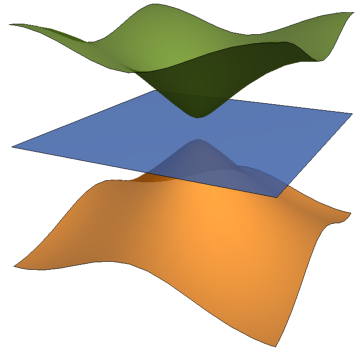}
    \caption{Gapped  ($\delta = 0.1$)}
    \label{subfigure:lieb_lattice_gapped_energy_dispersion}
  \end{subfigure}
  \caption{
    Lieb lattice and its band structure.
    (a) Lieb lattice.
    (b) The energy band for $\delta=0$ has a Dirac cone and a flat band.
    (c) $\delta$ generates a gap between the upper and lower bands.
  }
  \label{figure:lieb_lattice}
\end{figure}

One well-known example of flat-band generating lattices is the Lieb lattice, which has a band structure with a flat band together with a Dirac cone crossing the FB by its tip.
See figures
\ref{subfigure:lieb_lattice_geometric_structure}
and
\ref{subfigure:lieb_lattice_gapless_energy_dispersion}. 
Lieb lattice has been realized in photonic lattices
\cite{Guzman-Silva:2014:10.1088_1367-2630_16_6_063061, Mukherjee:2015:10.1103_PhysRevLett.114.245504, Vicencio:2015:10.1103_PhysRevLett.114.245503, Taie:2015:10.1126_sciadv.1500854, Xia:2016:10.1364_OL.41.001435},
on metal substrates
\cite{Slot:2017:10.1038_nphys4105, Drost:2017:10.1038_nphys4080},
and in organic frameworks
\cite{Cui:2020:10.1038_s41467-019-13794-y},
while its inorganic realization is still awaiting.  

The flat band has infinitely strong coupling because the effective coupling can be defined as the ratio of the potential and the kinetic energies and the latter is quenched.
Therefore the holographic description
\cite{Maldacena:1999:10.1023_A:1026654312961, Witten:1998:ATMP.1998.v2.n2.a2, Gubser:1998:10.1016_S0370-2693(98)00377-3, Hartnoll:2009:10.1088_0264-9381_26_22_224002, Herzog:2009:10.1088_1751-8113_42_34_343001, McGreevy:2010:10.1155_2010_723105, Horowitz:2011:10.1007_978-3-642-04864-7_10, Sachdev:2011:10.1007_978-3-642-04864-7_9}
of it would be very interesting.
There have been many interesting spectral functions of fermions
\cite{Lee:2009:10.1103_PhysRevD.79.086006, Liu:2011:10.1103_PhysRevD.83.065029, Cubrovic:1993:10.1126_science.1174962, Faulkner:2011:10.1098_rsta.2010.0354, Oh:2021:10.1007_JHEP01(2021)053, Liu:2018:10.1007_JHEP12(2018)072, Seo:2018:10.1007_JHEP08(2018)077, Chakrabarti:2019:10.1007_JHEP07(2019)037, Oh:2021:10.1007_JHEP11(2021)207}
and one particularly interesting one is the work by Laia and Tong
\cite{Laia:2011:10.1007_JHEP11(2011)125}
where it was shown that a flat band can be realized if one chooses a particular boundary action.

The ideal flat band model should have a flat that is band well separated from other bands.
Only in such case, observed phenomena can be safely attributed to the flat band.
The flat-band electrons move much more slowly than those in other bands so that, when two bands coexist, the effect of the flat band would not be clear.  
In order to get a well isolated flat band from the Lieb lattice band, we need to gap out the Dirac band. 

In this paper, we first point out that the  model of ref. \cite{Laia:2011:10.1007_JHEP11(2011)125}  actually describes an analog of the Lieb lattice which has a flat band crossed by a Dirac band.
In this holographic realization, however, we emphasize that the Dirac band has a branch cut singularity rather than a pole, unlike weakly interacting systems.
It can be interpreted as a consequence of the spectral transfer from the Dirac band to the flat band.
Since such a transfer is one of the characteristic phenomena of strong correlation, it is interesting by itself.
However, it also means that the flat band in the model does not  correspond to a band that is well separated from other bands.   

We will  introduce   an exactly solvable holographic model  where the flat band is well separated from the other bands by gapping the Dirac band. 
We identify the gap generating parameter of the holographic model in terms of the one in the lattice model by comparing the two models. 

We also find a few other ways to quantize the holographic fermions by finding new boundary actions.
We express the Green functions in the new  quantizations in terms of those  in the standard quantization.
 
\section{Realizing flat bands in lattice and holographic models}
\subsection{The tight-binding model of the Lieb lattice}
The tight-binding Hamiltonian of the Lieb lattice is given by
\cite{Julku:2016:10.1103_PhysRevLett.117.045303}
\begin{equation}
  \small
  \begin{split}
    H = &\ \sum_{\vec{R}\in  Lattice} [(1 + \delta) (c_{\vec{R}, B}^{\dagger} c_{\vec{R}, A} + c_{\vec{R}, C}^{\dagger} c_{\vec{R}, B}) + (1 - \delta) (c_{\vec{R} + \hat{x}, B}^{\dagger} c_{\vec{R}, A} + c_{\vec{R} - \hat{y}, C}^{\dagger} c_{\vec{R}, B}) + \text{h.c.}] \cr
          = &\ \int_{BZ} d^{2}{\vec{k}} \{ [(1 + \delta) + (1 - \delta) e^{i k_{x}}] c_{\vec{k}, A}^{\dagger} c_{\vec{k}, B} + [(1 + \delta) + (1 - \delta) e^{-i k_{y}}] c_{\vec{k}, B}^{\dagger} c_{\vec{k}, C} + \text{h.c.} \},
  \end{split}
\end{equation}
where the $A$, $B$, and $C$ label the atoms in a unit cell enclosed by the cyan dashed line in figure
\ref{subfigure:lieb_lattice_geometric_structure}.
The lattice constant $a$ is set to be 1 and 
  $\delta$ is the staggered hopping parameter.
Notice that the second line comes by the Fourier transform
$c_{\vec{R},j} =\int_{BZ}c_{ {\vec{k}},j}e^{i\vec{k}\cdot\vec{R}} $ with $j=A,B,C$.

Considering the nearest neighbor hoppings only, it is not difficult to show that the dispersion relation of the above tight-binding model is given by
\begin{equation}
  \epsilon_{\vec{k}} = 0, \pm 2 \sqrt{(1 + \delta^{2}) + (1 - \delta^{2}) (\cos k_{x} + \cos k_{y}) / 2},
\end{equation}
and the gap between the upper and lower bands at the $k_{x} = k_{y} = \pi$ ($M$-point) is
\begin{equation}
  \Delta_{\text{gap}} /2= 2 \sqrt{2} |\delta|.
\end{equation}
The band structures of this model for gapless and gapped cases are plotted in figures
\ref{subfigure:lieb_lattice_gapless_energy_dispersion}
and
\ref{subfigure:lieb_lattice_gapped_energy_dispersion},
respectively.

One should notice that the sum over the discrete lattice is replaced by the integral over the momentum over the Brillouin zone, which should be replaced by the infinite momentum space when we take the low energy limit.
In this limit, {\it realizing a lattice is reduced to realizing the band structure  near the $\Gamma$ point.}
This is the sense of introducing the lattice structure in holographic theory.
After the obvious scale in which lattice constants are taken to be zero so that the Brillouin zone becomes infinite, the other scales are generated by the interaction terms.
In the previous work of some of the authors, a few interesting band structures were generated by considering the interaction term of the form $\Phi_{A}{\bar\psi}\Gamma^{A}\psi$ with symmetry breaking condensation of $\Phi_{A}$ field
\cite{Oh:2021:10.1007_JHEP01(2021)053}.

\subsection{Holographic flat band models}
The fermion action is given by the sum of a bulk action and a boundary action $S_{\text{bulk}}+ S_{\text{bdy}}$, where
\begin{equation}
  S_{\text{bulk}} = \int_{\text{bulk}} \dd^{4} x \sqrt{-g} i \bar{\psi} \left[ \frac{1}{2} (\overrightarrow{\slashed{D}} - \overleftarrow{\slashed{D}}) - m - \Phi \right] \psi ,
\end{equation}
\begin{equation}
  \dd s^{2} = g_{A B} \dd x^{A} \dd x^{B} = \frac{1}{u^{2}} (-f \dd t^{2} + \dd x^{2} + \dd y^{2}) + \frac{\dd u^{2}}{f u^{2}},
\end{equation}
\begin{equation}
  \bar{\psi} = \psi^{\dagger} \Gamma^{\underbar{t}}, \quad
  \overrightarrow{\slashed{D}} = \Gamma^{a} \tensor{e}{_{a}^{B}} (\partial_{B} + \tfrac{1}{4} \omega_{B c d} \Gamma^{c d} - i q A_{B}), \label{DAt}
\end{equation}
\begin{equation}
  f = 1 - \left( 1 + \frac{1}{4} u_{h}^{2} \mu^{2} \right) \left( \frac{u}{u_{h}} \right)^{3} + \frac{1}{4} u_{h}^{2} \mu^{2} \left( \frac{u}{u_{h}} \right)^{4},
\end{equation}
\begin{equation}
  A_{t} = \mu \left( 1 - \frac{u}{u_{h}} \right), \quad
  T = \frac{1}{4 \pi u_{h}} \left( 3 - \frac{1}{4} u_{h}^{2} \mu^{2} \right).
\end{equation}
The indices $A, B, \cdots$ are for the bulk spacetime denoting $(t, x, y, u)$, and $a, b, \cdots$ are the tangent space indices denoting $(\underbar{t}, \underbar{x}, \underbar{y}, \underbar{u})$.
For simplicity, we take the background metric fixed and we consider the configuration of the scalar field where only the source term is present so that an analytic result is allowed.
Our gamma matrices are
\begin{align}
  \Gamma^{\underbar{t}} = &\ \sigma_{1} \otimes i \sigma_{2} = \mqty(0 & i \sigma_{2} \\ i \sigma_{2} & 0), &
  \Gamma^{\underbar{x}} = &\ \sigma_{1} \otimes \sigma_{1} = \mqty(0 & \sigma_{1} \\ \sigma_{1} & 0), \\
  \Gamma^{\underbar{y}} = &\ \sigma_{1} \otimes \sigma_{3} = \mqty(0 & \sigma_{3} \\ \sigma_{3} & 0), &
  \Gamma^{\underbar{u}} = &\ \sigma_{3} \otimes \sigma_{0} = \mqty(\sigma_{0} & 0 \\ 0 & -\sigma_{0}), \\
  \Gamma^{5} = &\ i \Gamma^{\underbar{t}} \Gamma^{\underbar{x}} \Gamma^{\underbar{y}} \Gamma^{\underbar{u}}, &
  \Gamma^{a b} = &\ \tfrac{1}{2} \comm*{\Gamma^{a}}{\Gamma^{b}}.
\end{align}
Taking the derivative of the bulk action, we have
\begin{equation}
  \delta S_{\text{bulk}} = (\text{EOM term}) + \frac{1}{2} \int_{\text{bdy}} \dd^{3} x \sqrt{-h} (\bar{\psi} i \Gamma^{\underbar{u}} \delta \psi - \delta \bar{\psi} i \Gamma^{\underbar{u}} \psi), \label{variation}
\end{equation}
where $h = g g^{u u}$ and the bulk Dirac equation is given by
\begin{equation}
  (\overrightarrow{\slashed{D}} - m - \Phi) \psi = 0.
\end{equation}
To simply deal with the equation of motion, we will take an ansatz that
\begin{equation}
  \psi(t, x, y, u) = (-h)^{-1/ 4} e^{-i \omega t + i k_{x} x + i k_{y} y} \phi(u).
\end{equation}

The boundary action should be chosen such that its variation kills the unwanted degrees of freedom in the second term of eq. (\ref{variation}) so that the equation of motion of the wanted degrees of freedom can be granted, which significantly restricts the possibilities.
For detail, see the appendix.
It turns out that to get the flat band, the choice of the boundary action is the key \cite{Laia:2011:10.1007_JHEP11(2011)125}. 
When we take a boundary action of the form
\begin{equation}
  S_{\text{bdy}} = \frac{1}{2} \int_{\text{bdy}} \dd^{3} x \sqrt{-h} \bar{\psi} \Gamma\psi
,\end{equation}
  the spectral function has a flat band for $\Gamma=\Gamma^{\underbar{x} \underbar{y}}$ \cite{Laia:2011:10.1007_JHEP11(2011)125}. 
We find that there are two more possible choices $ \Gamma =\pm \Gamma^{5 \underbar{x}} $ so that available  quantizations  are as follows:
\be
 \Gamma = \pm i \mathbb{I}_{4}, \quad \pm \Gamma^{\underbar{x} \underbar{y}} , \quad  \pm \Gamma^{5 \underbar{x}}.
\ee
The role of these boundary actions is to project out half of the degrees of freedom and the first choices $\pm i \mathbb{I}_{4}$ have been called standard ($+$ sign) and alternative ($-$ sign) quantization respectively, and the second two choices are called mixed quantization without distinguishing the signs because the sign difference does not make any essential difference for the mixed quantization.
For the third ones, they are new and we call them chiral quantization where the sign change in the boundary action  gives a sign flip of $k_{y}$ in the final Green function.
We will see that flat bands exist only for the mixed quantization with $S_{bdy}\sim {\bar\psi}\Gamma^{xy}\psi$.

\subsection{Green functions for various boundary actions}\label{appendix:changing_boundary_actions}
In this subsection, we consider the general problem of possible boundary actions that can project specific two components which are the boundary degrees of freedom out of four bulk components.
Then the boundary Green functions are calculated accordingly.

Taking the derivative of the bulk action, we have
\begin{gather}
  \delta S_{\text{bulk}} = (\text{EOM term}) + \frac{1}{2} \int_{\text{bdy}} \dd^{3} x \sqrt{-h} (\bar{\psi} i \Gamma^{\underbar{u}} \delta \psi - \delta \bar{\psi} i \Gamma^{\underbar{u}} \psi), \\
  h = g g^{u u}, \quad (\overrightarrow{\slashed{D}} - m - \Phi) \psi = 0.
\end{gather}
The near-boundary solution of the equation of motion is given by \cite{Iqbal:2009:10.1002_prop.200900057}
\begin{align}
  \psi = (-h)^{-1 / 4} e^{-i \omega t + i \vec{k} \cdot \vec{x}} \phi, & &  {\phi}  \simeq \mqty(a_{1} u^{-m} + b_{1} u^{m + 1} \\ a_{2} u^{-m} + b_{2} u^{m + 1} \\ c_{1} u^{-m + 1} + d_{1} u^{m} \\ c_{2} u^{-m + 1} + d_{2} u^{m}).
\end{align}
When we take a boundary action of the form
\begin{equation}
  S_{\text{bdy}} = \frac{1}{2} \int_{\text{bdy}} \dd^{3} x \sqrt{-h} \bar{\psi} \Gamma \psi = \frac{1}{2} \int_{\text{bdy}} \dd^{3} x \bar{\phi} \Gamma \phi,
\end{equation}
then we have
\begin{equation}
  \begin{split}
    \delta S_{\text{tot}} = &\ \frac{1}{2} \int_{\text{bdy}} \dd^{3} x [\bar{\phi} (\Gamma + i \Gamma^{\underbar{u}}) (\delta \phi) + (\delta \bar{\phi}) (\Gamma - i \Gamma^{\underbar{u}}) \phi] \\
    := &\ \int_{\text{bdy}} \dd^{3} x [\bar{\phi} \Gamma_{+} (\delta \phi) + (\delta \bar{\phi}) \Gamma_{-} \phi],
  \end{split}
  \label{varStot}
\end{equation}
where $\bar{\phi} := \phi^{\dagger} \Gamma^{\underbar{t}}$ and $\Gamma_{\pm} := \frac{1}{2} (\Gamma \pm i \Gamma^{\underbar{u}})$.
For the number of $\delta \psi_{i}$'s to be reduced from $4$ to $2$, we need the following conditions for the boundary actions:
\begin{equation}
  \rank \Gamma_{\pm} = 2 \Rightarrow \rank (\Gamma \pm i \Gamma^{\underbar{u}}) = 2,
\end{equation}
\begin{equation}
  [\bar{\phi} \Gamma_{+} (\delta \phi)]^{\dagger} = (\delta \bar{\phi}) \Gamma_{-} \phi
  \Rightarrow (\Gamma^{\underbar{t}} \Gamma_{+})^{\dagger} = \Gamma^{\underbar{t}} \Gamma_{-}
  \Rightarrow \Gamma = \Gamma^{\underbar{t}} \Gamma^{\dagger} \Gamma^{\underbar{t}}
  \Rightarrow \Gamma^{\dagger} = \Gamma^{\underbar{t}} \Gamma \Gamma^{\underbar{t}}. \label{gad}
\end{equation}

Substituting the near-boundary solution to the total action, for $0 < |m| < \frac{1}{2}$,
\begin{equation}
  \begin{split}
    S_{\text{tot}} = &\ S_{\text{bdy}} = \frac{1}{2} \int_{\text{bdy}} \dd^{3} x \bar{\phi} \Gamma \phi \\
    := &\ \frac{1}{2} \int_{\text{bdy}} \dd^{3} x \overline{\mqty(A u^{-m} + B u^{m + 1} \\ C u^{-m + 1} + D u^{m})} \mqty(\gamma_{1 1} & \gamma_{1 2} \\ \gamma_{2 1} & \gamma_{2 2}) \mqty(A u^{-m} + B u^{m + 1} \\ C u^{-m + 1} + D u^{m}) \\
    = &\ \frac{1}{2} \int_{\text{bdy: $u\to 0$}} \dd^{3} x (\bar{A} \gamma_{2 1} A u^{-2 m} + \bar{A} \gamma_{2 2} D + \bar{D} \gamma_{1 1} A + \bar{D} \gamma_{1 2} D u^{2 m}) + (\text{vanishing terms}),
  \end{split}
\end{equation}
where $\bar{A} := A^{\dagger} i \sigma_{2}$ and $\bar{D} := D^{\dagger} i \sigma_{2}$.
For the total action to converge,  $\gamma_{1 2} = 0$ if ${-\frac{1}{2} < m < 0}$,  while $\gamma_{2 1} = 0$ if ${0 < m < \frac{1}{2}}$. 
%
In any case, if we define a 4-component spinor by $\phi_{0} := \mqty(A \\ D)$ and $\bar{\phi}_{0} := \phi_{0}^{\dagger} \Gamma^{\underbar{t}}$, then finite part of the total action can be written as
\begin{equation}
  S_{\text{tot, on-shell}} = \frac{1}{2} \int_{\text{bdy}} \dd^{3} x (\bar{A} \gamma_{2 2} D + \bar{D} \gamma_{1 1} A) = \frac{1}{2} \int_{\text{bdy}} \dd^{3} x {\bar \phi_{0}}\mqty(\gamma_{1 1} & 0 \\ 0 & \gamma_{2 2}) \phi_0.
\end{equation} 
Similarly, the finite part of $\delta S_{tot}$ in eq. (\ref{varStot}) can be written as
\begin{equation}
  \begin{split}
    \delta S_{\text{tot}} = \int_{\text{bdy}} \dd^{3} x &\ \left\{ \bar{\phi}_{0}  \left[ \frac{1}{2} \left( \mqty(\gamma_{1 1} & 0 \\ 0 & \gamma_{2 2}) + i \Gamma^{\underbar{u}} \right) \right] (\delta \phi_0) \right. \\
    &\ \qquad \left. + (\delta {\bar\phi}_0) \left[ \frac{1}{2}  \left( \mqty(\gamma_{1 1} & 0 \\ 0 & \gamma_{2 2}) - i \Gamma^{\underbar{u}} \right) \right] \phi_0 \right\}.
  \end{split}
\end{equation}
%
Therefore, without loss of generality, we can assume that  $\Gamma$  has to be in the diagonal form: 
\begin{equation}
  \Gamma = \mqty(\gamma_{1 1} & 0 \\ 0 & \gamma_{2 2}).
\end{equation}
For such $\Gamma$'s, the total action converges for all $-\frac{1}{2} < m < \frac{1}{2}$, and we have
\begin{equation}
  S_{\text{tot}} = \frac{1}{2} \int_{\text{bdy}} \dd^{3} x {\bar \phi_0} \Gamma \phi_0 = \int_{\text{bdy}} \dd^{3} x (\bar{\phi}_{0} \Gamma_{+} \phi_{0} + \bar{\phi}_{0} \Gamma^{\underbar{t}} \Gamma_{+}^{\dagger} \Gamma^{\underbar{t}} \phi_{0}),
\end{equation}
\begin{equation}
  \begin{split}
    \delta S_{\text{tot}} = &\ \int_{\text{bdy}} \dd^{3} x \left\{ \bar{\phi}_{0} \left[ \frac{1}{2} (\Gamma + i \Gamma^{\underbar{u}}) \right] (\delta \phi_0) + (\delta \bar{\phi}_{0}) \left[ \frac{1}{2} (\Gamma - i \Gamma^{\underbar{u}}) \right] \phi_{0} \right\} \\
    = &\ \int_{\text{bdy}} \dd^{3} x [\bar{\phi}_{0} \Gamma_{+} (\delta \phi_0) + (\delta \bar{\phi}_{0}) \Gamma^{\underbar{t}} \Gamma_{+}^{\dagger} \Gamma^{\underbar{t}} \phi_0].
  \end{split}
\end{equation}
Here, we assumed that $\Gamma^{\underbar{t}} \Gamma$ is Hermitian so that $\delta S_{\text{tot}}$ is real. This is the reason for eq. (\ref{gad}).
%

If we restrict ourselves to  simple boundary actions with $\Gamma \propto \mathbb{I}_{4}, \Gamma^{5}, \Gamma^{a}, \Gamma^{5 a}, \Gamma^{a b}$, then there are eight  choices 
\be
\Gamma = \pm i \mathbb{I}_{4}, \pm \Gamma^{5 \underbar{x}}, \pm \Gamma^{5 \underbar{y}}, \pm \Gamma^{\underbar{x} \underbar{y}}\label{230}\ee
that satisfies the conditions mentioned above.
For each of above $\Gamma$'s, we have $4 \times 4$ projection operator $\mathbb{P}$:
\begin{equation}
  \mathbb{P} := -i \Gamma^{\underbar{u}} \Gamma_{+} \quad \Rightarrow \quad \mathbb{P}^{2} = \mathbb{P}, \quad \mathbb{P}^{\dagger} = \mathbb{P}.
\end{equation}
Defining $\mathbb{J} := \mathbb{P} \phi_{0}$ and $\mathbb{C} := i \mathbb{P} \Gamma^{\underbar{u}} \Gamma^{\underbar{t}} \phi_{0}$, we obtain
\begin{equation}
  \begin{split}
    \delta S_{\text{tot}} = &\ \int_{\text{bdy}} \dd^{3} x [\bar{\phi}_{0} (i \Gamma^{\underbar{u}} \mathbb{P}) \mathbb{P} (\delta \phi_{0}) + (\delta \bar{\phi}_{0}) \Gamma^{\underbar{t}} \mathbb{P} (i \Gamma^{\underbar{u}} \mathbb{P})^{\dagger} \Gamma^{\underbar{t}} \phi_{0}] \\
    = &\ \int_{\text{bdy}} \dd^{3} x [(i \mathbb{P} \Gamma^{\underbar{u}} \Gamma^{\underbar{t}} \phi_{0})^{\dagger} (\mathbb{P} \delta \phi_{0}) + (\mathbb{P} \delta \phi_{0})^{\dagger} (i \mathbb{P} \Gamma^{\underbar{u}} \Gamma^{\underbar{t}} \phi_{0})]\\
    = &\ \int_{\text{bdy}} \dd^{3} x [\mathbb{C}^{\dagger} (\delta \mathbb{J}) + (\delta \mathbb{J}^{\dagger}) \mathbb{C}],
  \end{split}
\end{equation}
\begin{equation}
  S_{\text{tot}} = \int_{\text{bdy}} \dd^{3} x (\mathbb{C}^{\dagger} \mathbb{J} + \mathbb{J}^{\dagger} \mathbb{C}).
\end{equation}
As one can see from the table \ref{table:p_p_and_q_matrices_table}, $\mathbb{J}$ has only two independent components.
Therefore we want to write $\mathbb{J}$ and $\mathbb{C}$ as 2-component spinors, which we call $J$ and $C$ respectively.
We try to write
\begin{equation}
  \Gamma^{\underbar{t}} \Gamma_{+} = i \Gamma^{\underbar{t}} \Gamma^{\underbar{u}} \mathbb{P} = Q^{\dagger} P,
\end{equation}
where $Q$ and $P$ are $2 \times 4$ matrices.  Such choice for each $\Gamma$ is given in table \ref{table:p_p_and_q_matrices_table}.
Then, we have
\begin{align}
  \delta S_{\text{tot}} = &\ \int_{\text{bdy}} \dd^{3} x [(P \delta \phi_0)^{\dagger} (Q \phi_0) + \text{h.c.}] = \int_{\text{bdy}} \dd^{3} x [(\delta J^{\dagger}) C + \text{h.c.}], \label{231}\\
  S_{\text{tot}} = &\ \frac{1}{2} \int_{\text{bdy}} \dd^{3} x (J^{\dagger} C + \text{h.c.}) = \frac{1}{2} \int_{\text{bdy}} \dd^{3} x (J^{\dagger} G J + \text{h.c.}),
\end{align}
where $ C = Q \phi_0$, $J = P \phi_0$, and $C = G J$ so that 
\begin{equation}
  Q \phi_0 = G P \phi_0. \label{234}
\end{equation}
Now, $J$ should be free to choose at the boundary so that we can take the derivative with respect to it.
Taking the second derivative of the total on-shell action with respect to $J$, we see that $G$ is the Green function.

\begin{table}[t]
  \centering
  \includegraphics[width = 6 in]{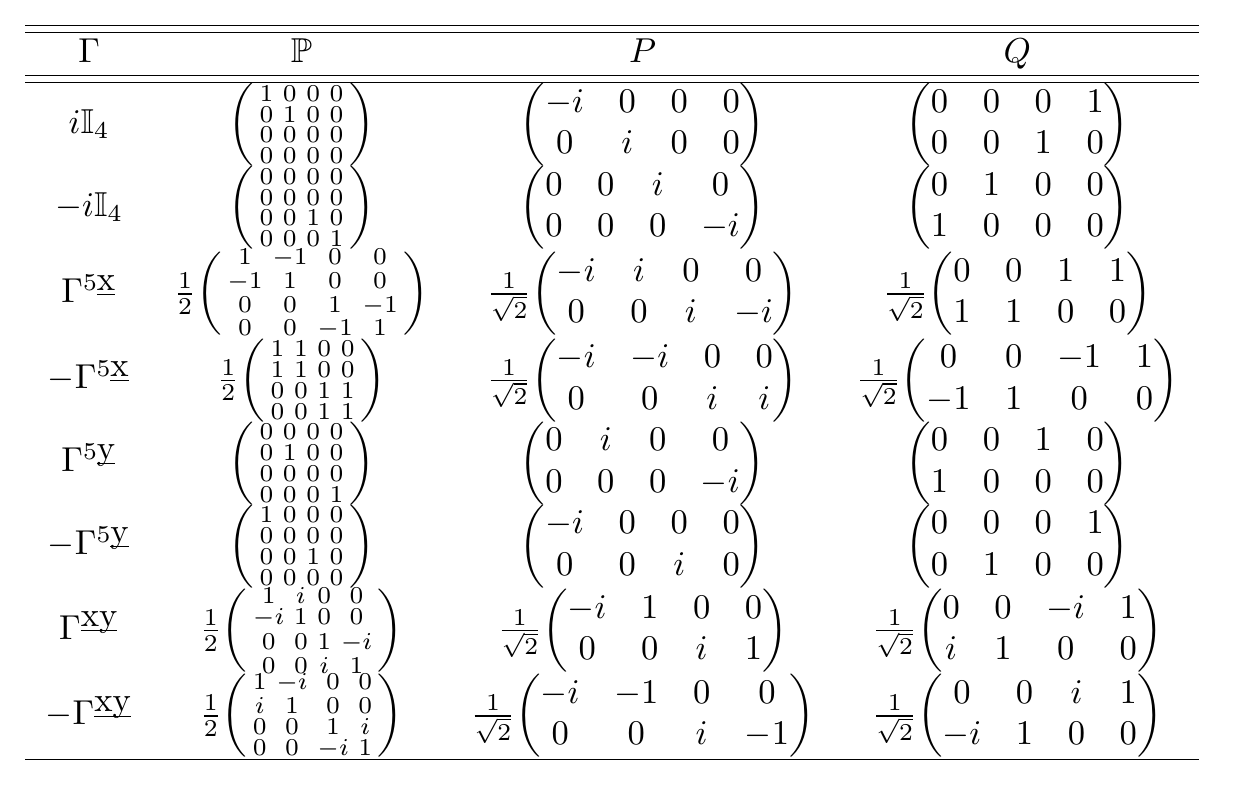}
  \caption{$\mathbb{P}$, $P$, and $Q$ matrices table}
  \label{table:p_p_and_q_matrices_table}
\end{table}

Notice that the decomposition of $i \Gamma^{\underbar{t}} \Gamma^{\underbar{u}} \mathbb{P}$ into $P$ and $Q$ is not unique by a $2\times 2$ unitary matrix $U: \ P\to UP, \quad Q \to U Q$.
However, from eq. (\ref{231}), it is obvious that $G$ does not depend on such a choice of $U$.
For the eight choices listed in eq. (\ref{230}), we tabulate the solutions for $P$ and $Q$ in table \ref{table:p_p_and_q_matrices_table}.
For the standard quantization $\Gamma = i \mathbb{I}_{4}$, eq. (\ref{234}) together with $Q=\mqty(0 & \sigma_{1})$ and $P= \mqty(-i \sigma_{3} & 0) $
gives 
\be
D = -i \sigma_{1} G^{S} \sigma_{3} A, \label{DA}
\ee
where the superscript $S$ represents the standard quantization.
Therefore, when we know $G^{S}$, we can directly find  $G$'s for other quantizations by using eqs. (\ref{234}) and (\ref{DA}).
That is, $\ Q \mqty(\sigma_{0} \\ -i \sigma_{1} G^{S} \sigma_{3}) A = G P \mqty(\sigma_{0} \\ -i \sigma_{1} G^{S} \sigma_{3}) A$ gives
 \begin{equation}
  \begin{split}
    G = \left[ Q \mqty(\sigma_{0} \\ -i \sigma_{1} G^{S} \sigma_{3}) \right] \left[ P \mqty(\sigma_{0} \\ -i \sigma_{1} G^{S} \sigma_{3}) \right]^{-1}.
  \label{equation:connection_between_quantizations}
  \end{split}
\end{equation}
 The necessary $P$ and $Q$ data are given in table \ref{table:p_p_and_q_matrices_table}.
%
For example, the Green functions $G^{\Gamma}$ for $\Gamma = \Gamma^{5 \underbar{x}}, \Gamma^{\underbar{x} \underbar{y}}$ and $G^{A}$ ( where $A$  for the alternative quantization) can be calculated by the following:
\bea
G^{A} 
  &&= \frac{-1}{\det G^{S}} \mqty( G_{1 1}^{S} & G_{2 1}^{S} \\ G_{1 2}^{S} & G_{2 2}^{S}) 
  = -\frac{(G^{S})^{T}}{\det G^{S}},\quad \quad\\
   {\tr G}^{ {xy}} &&= \frac{2 \det G^{S} - 2}{\tr G^{S} - \tr \sigma_{2} G^{S}},
  \quad  \quad
   {\tr G}^{ {5x}} = \frac{2 \det G^{S} - 2}{\tr G^{S} - \tr \sigma_{1} G^{S}},
   \label{GforM}
\eea
\bea
  G^{x y} &&= \frac{1}{\tr G^{S} - \tr \sigma_{2} G^{S}}
  \mqty(
  2 \det G^{S} & -i \tr \sigma_{1} G^{S} + \tr \sigma_{3} G^{S} \\
  i \tr \sigma_{1} G^{S} + \tr \sigma_{3} G^{S} & -2
  ), \label{GrinmixedQz}\\
  G^{5 x} &&= \frac{1}{\tr G^{S} - \tr \sigma_{1} G^{S}}
  \mqty(
  2 \det G^{S} & -\tr \sigma_{2} G^{S} + i \tr \sigma_{3} G^{S}
  \label{Gmixed}
   \\
  -\tr \sigma_{2} G^{S} - i \tr \sigma_{3} G^{S} & -2
  ).
\eea
If we set with $k_{y}=0$ in $G^{ {xy}}$, the result is consistently reduced to that of \cite{Li:2011:10.1007_JHEP11(2011)018}.
To study the shape of the Fermi sea or contour plot of the spectral function at the fixed $\omega$, we need the spectral function as a function of both $k_{x}$ and $k_{y}$, because there is no rotational symmetry. 
 \begin{figure}[t]
  \centering
  \begin{subfigure}[h]{0.3 \textwidth}
    \centering
    \includegraphics[width = 1.5 in]{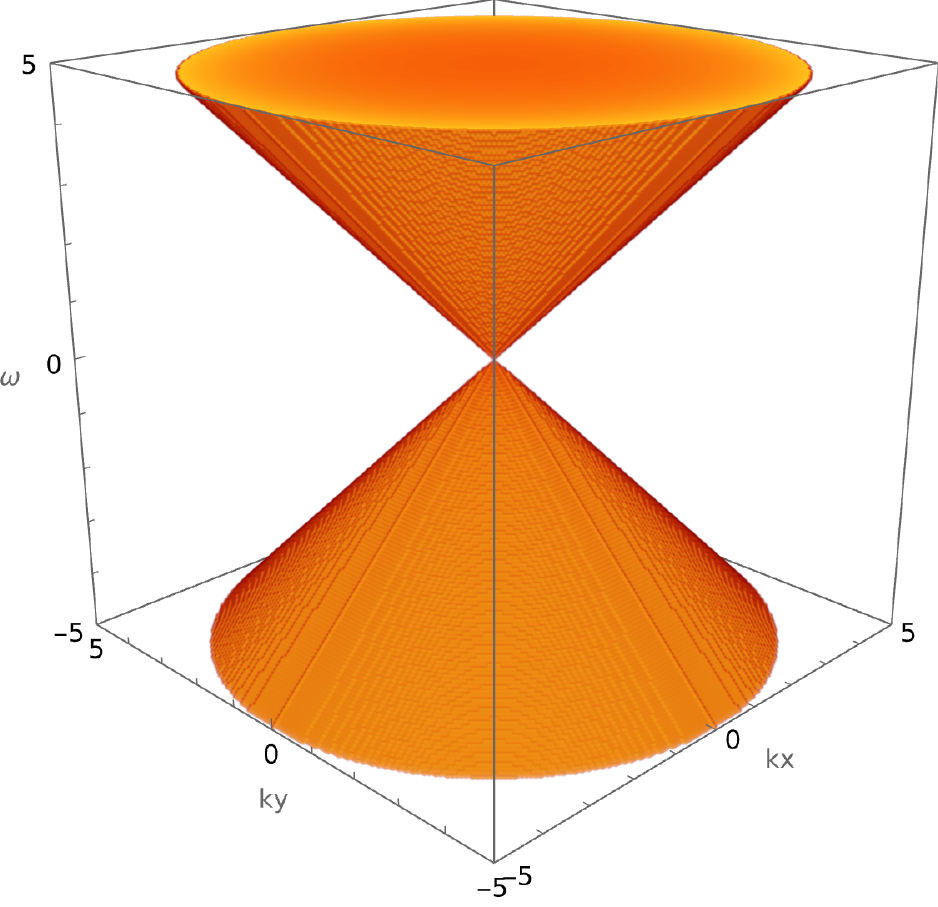}
    \caption{Standard ($\Gamma = i \mathbb{I}_{4}$)}
  \end{subfigure}
  \begin{subfigure}[h]{0.3 \textwidth}
    \centering
    \includegraphics[width = 1.5 in]{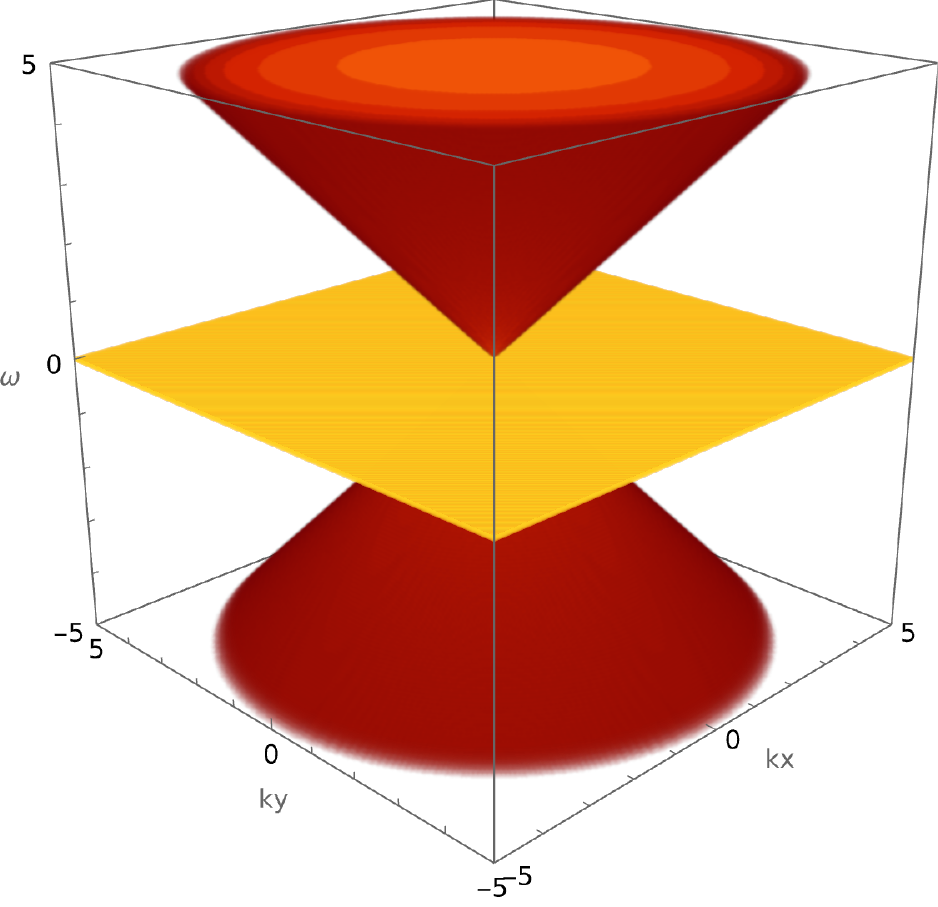}
    \caption{$\Gamma = \Gamma^{\underbar{x} \underbar{y}}$}
  \end{subfigure}
  \begin{subfigure}[h]{0.3 \textwidth}
    \centering
    \includegraphics[width = 1.5 in]{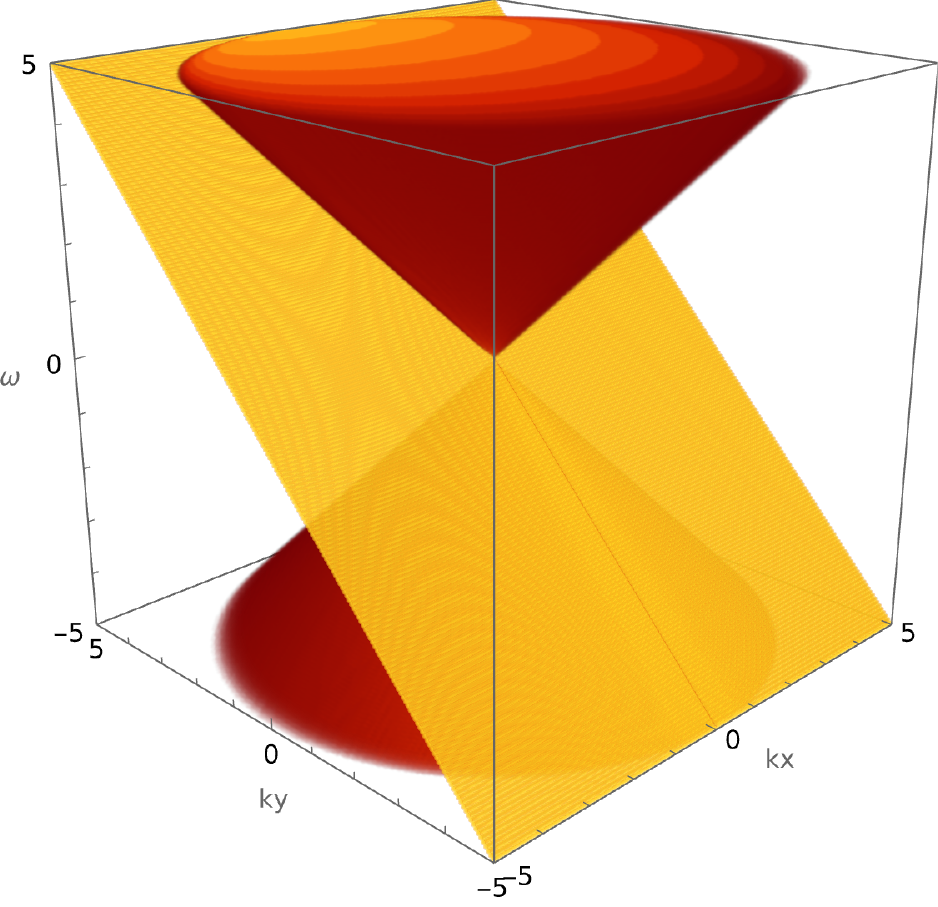}
    \caption{$\Gamma = \Gamma^{5 \underbar{x}}$}
  \end{subfigure}
  \caption{
    Holographic spectral functions for different boundary actions
    $S_{bdy}=  \int_{bdy} {\bar\psi} \Gamma \psi /2$ where
    $\Gamma  =   i \mathbb{I}_{4},  \Gamma^{\underbar{\it xy}}, \Gamma^{ \underbar{5\it j}}$ with $j=x,y$.
    $\pm S_{bdy}$ give essentially the same results with $k_{y}\leftrightarrow -k_{y}$.  }
\label{figure:spectral_functions_without_interaction}
\end{figure}
 
The spectral function is defined as the imaginary part of the retarded Green function:
\be 
  \rho^{A} = 2 \Im \tr G^{A}.
\ee
In figure \ref{figure:spectral_functions_without_interaction}, the spectral functions without scalar coupling are plotted for the zero bulk mass case.  Notice that as we mentioned in the introduction, the spectrum in the Dirac band is a branch cut   that is not a particle spectrum, which is a manifestation of strongly-coupled nature.
The spectrum for the $\Gamma^{\underbar{x} \underbar{\it y}}$ is very similar to the band structure of the Lieb lattice. In the  section 5, we describe the underlying mechanism for it. 

\section{Gapping the Dirac band in the Laia-Tong Model}

The model for gapping the Dirac band is simply given as the one with scalar coupling $\Phi\neq 0$.
The analytic result for the Green function in the standard quantization with general bulk fermion mass and non-vanishing scalar configurations was given in our previous work \cite{Oh:2021:10.1007_JHEP11(2021)207}:
For $\Phi=\Phi_{0}u$ with positive $\Phi_{0}$, 
\bea
G ^{S}&=&  \frac{  (4{\mu})^{\frac{1}{2}+m} 
\Gamma (-2 m)\Gamma \left(1+m+\nu\right)}{\left(k^2-w^2\right)
	\Gamma(-m+ \nu )\Gamma (1+2 m)} \gamma^\mu k_\mu \gamma^t ,
 \label{GRFM0}
\eea
 where parameters $\mu$ and $\nu$ are given by 
 \be
 \mu=k^2-w^2+\Phi_0^2, \quad   
\nu=\frac{ m\Phi_{0} }{\sqrt{\mu}}.
\ee
The poles of the Green function are given by those of the gamma function at the non-positive integers so that the massive spectrum can be read off as
\be
\omega^{2}-k^{2}=\Phi_{0}^{2}\left(1-  {m^{2}}/{(n+m+1)^{2}}\right),  n=0,1,2 \cdots .
\ee

For $m = 0$, the tower of the discrete spectrum reduces to a single particle spectrum $\omega^{2}-k^{2}=\Phi_{0}^{2}$, the Green function simplifies to 
 \begin{equation}
  G^{S} = \frac{1}{\Phi_{0} + \sqrt{-\omega^{2} + k_{x}^{2} + k_{y}^{2} + \Phi_{0}^{2}}} \mqty(\omega -k_{x} & k_{y} \\ k_{y} & \omega  + k_{x}), \label{GwithPhi}
\end{equation}
and we have
\begin{align}
   {\tr G}^{S} = &\ \frac{2 \omega}{\Phi_{0} + \sqrt{-\omega^{2} + k_{x}^{2} + k_{y}^{2} + \Phi_{0}^{2}}}, \\
   {\tr G}^{  {xy}} = &\ \frac{-2 \sqrt{-\omega^{2} + k_{x}^{2} + k_{y}^{2} + \Phi_{0}^{2}}}{\omega+i\epsilon}, \\
   {\tr G}^{    {5x}} = &\ \frac{-2 \sqrt{-\omega^{2} + k_{x}^{2} + k_{y}^{2} + \Phi_{0}^{2}}}{\omega - k_{y}}.
\end{align}
The result for the   case  $ \Gamma =\Gamma^{\underbar{xy}}$ with  $\Phi_{0}=0$ is consistently reduced to that of  \cite{Laia:2011:10.1007_JHEP11(2011)125}, checking the consistency of our calculation. 
Now notice that $\tr G^S \propto \omega$, i.e, the k-dependence is cancelled between 
$  G^S_{11}$ and $G^S_{22}$ as we promised above. 

\begin{figure}[t]
  \centering
  \begin{subfigure}[h]{0.3 \textwidth}
    \centering
    \includegraphics[height = 1.5 in]{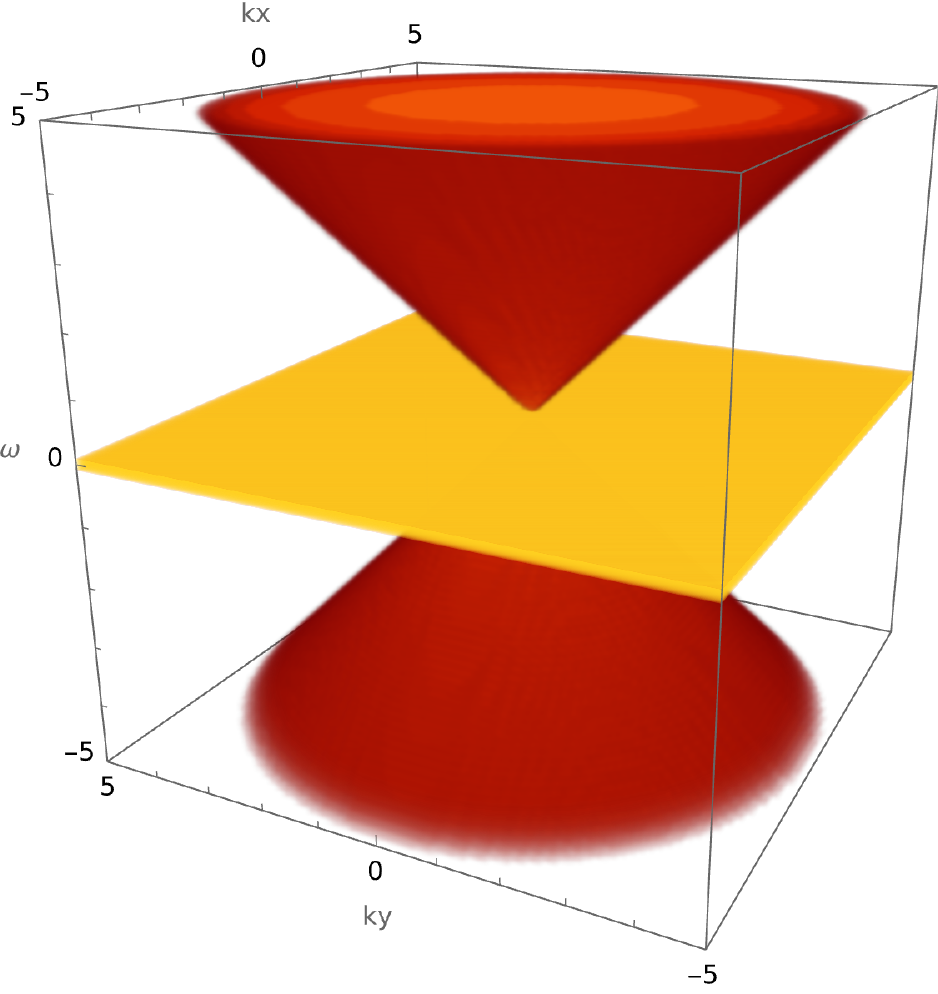}
    \caption{Gapless ($\Phi_{0} = 0$)}
  \end{subfigure}
  \begin{subfigure}[h]{0.3 \textwidth}
    \centering
    \includegraphics[height = 1.5 in]{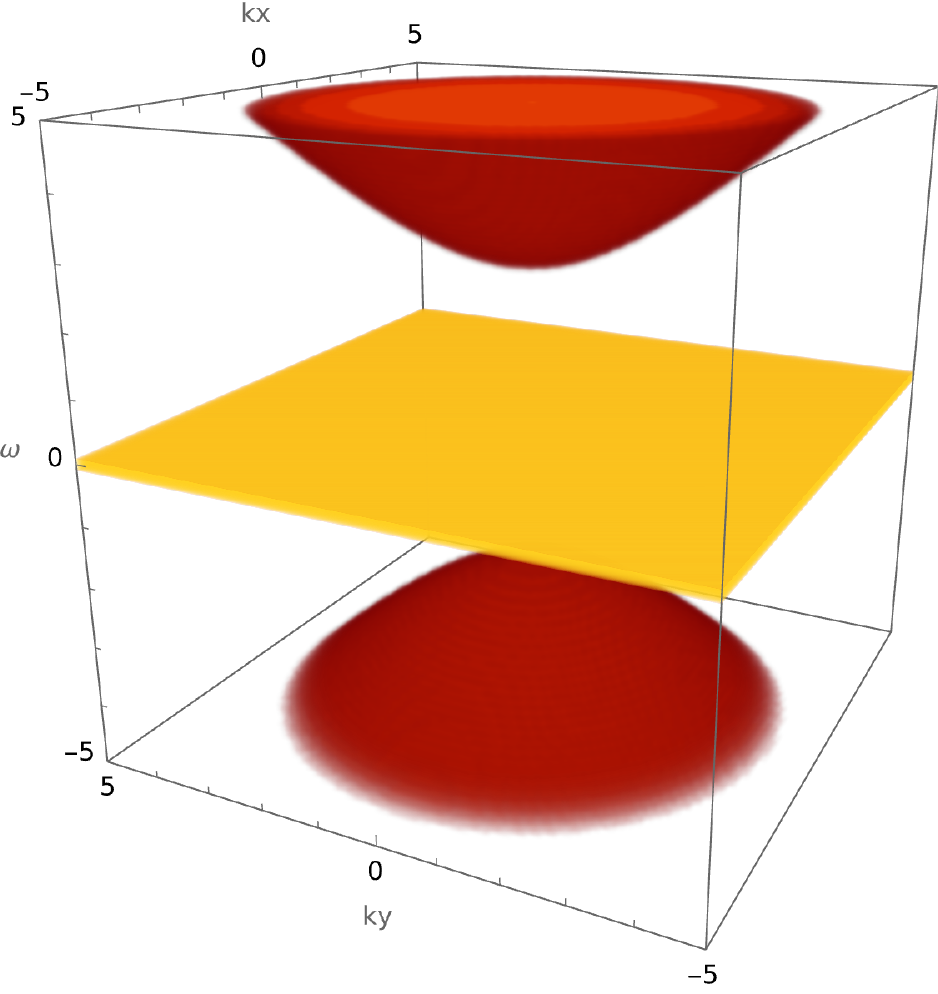}
    \caption{Gapped ($\Phi_{0} = 2.5$)}
  \end{subfigure}
  \caption{
    Holographic realization of Lieb lattice and its gapping.
    (a) The boundary action
    $\frac12 \int_{bdy} {\bar\psi}  \Gamma^{\underbar{\it xy}} \psi$
    without scalar coupling gives a flat band crossed by a Dirac band.
    (b) The $\Phi_{0}$ generates a gap between the upper and lower Dirac bands.
  } 
\label{figure:holographic_realization_of_lieb_lattice_flat_band}
\end{figure}

On the other hand, 
if we include  the scalar   field coupling, 
as we can see  
in figure \ref{figure:holographic_realization_of_lieb_lattice_flat_band},   a gap is generated between the upper and the lower bands just  like the case of the presence of staggered hopping parameter $\delta$ in the tight-binding model  shown in figure \ref{subfigure:lieb_lattice_gapped_energy_dispersion}.
By matching the gap size of the two models, we can identify 
\begin{equation}
  \frac{\Delta_{\text{gap}}}{2} = \underbrace{2 \sqrt{2} \,\delta}_{\text{Tight-Binding Model}} = \underbrace{\Phi_{0}}_{\text{Holographic Model}}.
\end{equation}
So that the turning on the scalar source $\Phi_{0}$ is equivalent to introducing alternation  in the lattice hoping, parametrized by $\delta$. 
 We remark that at the  finite temperature, the spectral weight is not delta function localized at the flat band, unlike the tight-binding model.

\section{ Laia-Tong model as the Holographic Lieb lattice}
 
\begin{figure}[t]
  \centering
  \begin{subfigure}[h]{0.3 \textwidth}
    \centering
    \includegraphics[width = 2 in]{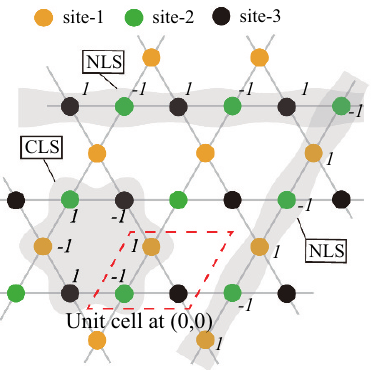}
    \caption{CLS in Kagome lattice}
  \end{subfigure}
  \hskip 2cm
 \begin{subfigure}[h]{0.3 \textwidth}
    \centering
    \includegraphics[width = 2 in]{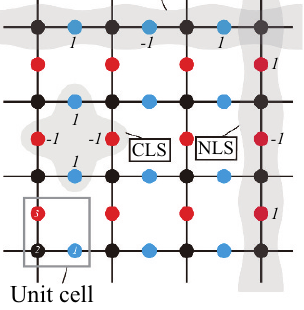}
    \caption{ CLS in Lieb Lattice}
  \end{subfigure}
   \caption{Compact localized states (CLS) are formed if two neighboring sites in the orbit contribute hopping amplitudes such that they cancel each other so that an electron can not escape the orbital. This figures are from ref. \cite{doi:10.1080/23746149.2021.1901606}.
  }
  \label{figure:CLS}
\end{figure}
In non-interacting or weakly interacting systems, it is easy to agree that two models with the same band structure can be identified.
This is because the Green function does not have features other than the locus of the poles, which is described by the delta function along the band dispersion.
In strongly interacting systems, the Green functions can have structures other than the locus of the singularity.
So  we look for further reasoning other than the spectral shape.  

The appearance of the Dirac band both holography as well as in the lattice is by now well known after the graphene.
 Perhaps the most serious mystery in the Laia-Tong model is the reason for the appearance of the the flat band, since the only feature of the model is the ``boundary condition'' (BC).
To see what is going on, we should understand the appearance of the flat band in lattice models.
A flat band means the electron does not move or  it is  localized in a restricted region.
It turns out that it can be understood as the cancellation of amplitudes coming from the neighboring sites so that electrons are confined in a small closed orbit in a lattice.
Such closed orbit is dubbed as the compact localized state (CLS) \cite{maimaiti2017compact}, which are generators of the flat band. In   figure \ref{figure:CLS}, 
we illustrate such CLS for the Lieb and Kagome lattices. 

What about the Laia-Tong model?
How can a "boundary condition"   give the flat band?
 For this, we should remind ourselves that the role of the boundary condition in the fermion holography is to select two degrees of freedom out of four spinor components in AdS$_4$ or AdS$_5$. The standard quantization   chooses the upper two components, while the alternative quantization select the lower two. In the Laia-Tong model, the so called mixed quantization chooses one from the upper two and one from the lower two such that their k-dependence in dispersion relation cancel precisely. 
Such cancellation structure is not so obvious in bare eye,  but the expression of Green functions in mixed quantization given in \eqref{GrinmixedQz} copied below shows such structure explicitly:
\bea
  G^{x y} = \frac{1}{\tr G^{S} - \tr \sigma_{2} G^{S}}
  \mqty(
  2 \det G^{S} & -i \tr \sigma_{1} G^{S} + \tr \sigma_{3} G^{S} \\
  i \tr \sigma_{1} G^{S} + \tr \sigma_{3} G^{S} & -2
  ).  \nonumber \eea  
Since $\tr \sigma_{2} G^{S}=0$, the poles of the Green function come from the zeroes of the $\tr G^{S}$ which is proportional to $\omega$.
That is, the flat band appears if 
\be
\tr{i\sigma_2 G^S}=0, \quad \hbox{ and } \tr{  G^S}=\omega\cdot (*).
\ee
The last identity can be explicitly demonstrated by the analytical form of the Green function given in eq. \eqref{GwithPhi} by $(\omega + k) + (\omega - k) = 2 \omega$.
 Therefore, in both lattice and holography models, the flat band comes from the cancellation between the amplitudes contributed from neighboring degrees of freedom.

Of course, just as in the non-interacting cases, 
the flat band alone does not characterize the given lattice.
The system is characterized only by the help of the accompanying non-flat band.
Therefore, different lattice models with flat band should be holographically realized by a particular interaction terms which correctly produces the accompanying non-flat band.

\section{Counting degree of freedom in holography}
\begin{figure}[t]
  \centering
  \begin{subfigure}[h]{0.3 \textwidth}
    \includegraphics[height = 1.7 in]{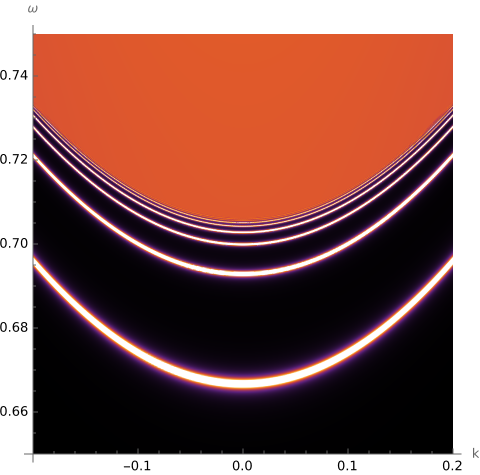}
    \caption{with source}
  \end{subfigure}
  \begin{subfigure}[h]{0.3 \textwidth}
    \includegraphics[height = 2.0 in]{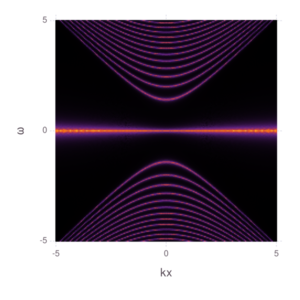}
    \caption{with condensation}
  \end{subfigure}
  \begin{subfigure}[h]{0.3 \textwidth}
    \includegraphics[height = 2.0 in]{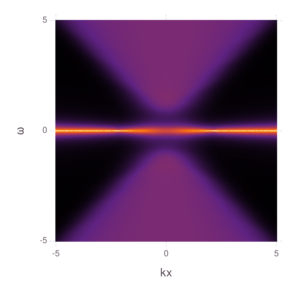}
    \caption{high $T$ with condensation}
    \label{subfigure:high_T_with_condensation}
  \end{subfigure}

  \caption{Tower of poles.
    The spectral density function in the mixed quantization near $k = 0$, $\omega = 0.7$ in the case of $m = -\frac{1}{4}$, $\Phi_{0}^{2} = \frac{1}{2}$ shows a ``tower of poles''.
  }
  \label{figure:tower_of_poles}
\end{figure}

Before we conclude, we want to draw some attention to the counting degrees of freedom (DOF) in holography which is a bit subtle but important.
The subtlety is that, while each band in the tight-binding theory contributes one DOF, the number of the bands is not necessarily the same as the DOF in holographic theory.
For example, in figure \ref{figure:tower_of_poles}, one can see that there are infinitely many bands, and we can show that, for zero temperature, they correspond to the simple poles of the gamma functions with non-integer residues.
This is clear from our eq. (\ref{GRFM0}) for the source and its sister expression in ref. \cite{Oh:2021:10.1007_JHEP11(2021)207} for the condensation.
Such an infinite tower of bands has been known in the early literature.
See \cite{Karch:2006pv,Allais:2012ye} for example.
It certainly reveals the presence of a new type of DOF.
So the problem exists not only in this ``holographic Lieb lattice model'' but also in wide classes of holographic theories.
The Dirac band in holography with a scalar field is also the sum of the infinite tower at $T=0$.
At finite temperature/density, however, these poles easily melt into a fuzzy single band.
See figure \ref{subfigure:high_T_with_condensation}.
Such melting happen for vanishing bulk fermion mass even near zero temperature, as shown in figures \ref{figure:TMeffect} and \ref{figure:effects_of_chemical_potential_and_temperature_in_mixed_quantization}.
Our fuzzy Dirac band is precisely such a case.

Although this looks surprising from the tight-binding point of view, it is not really a mystery.
Even in weakly interacting theory, if we include inter-band transition, the spectral weights can transfer from one band to the other.
Then each band's contribution to the degree of freedom is not necessarily fixed to be one, although their sum is fixed.
Now, for strongly interacting theory, a macroscopic number of particles can be entangled, and counting the degrees of freedom based on particle number in a unit cell loses its ground because entanglement and correlation can reach far beyond one cell.
From this point of view, the appearance of infinitely many bands is not surprising.

\section{The effects of charge density and temperature}

\begin{figure}[t]
  \centering
  \begin{subfigure}[h]{0.3 \textwidth}
    \centering
    \includegraphics[height =2 in]{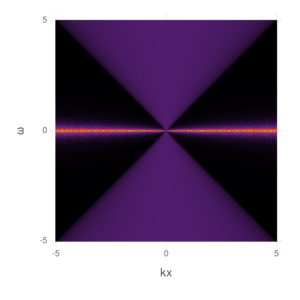}
    \caption{$\Phi = 0$, $\mu = 0$, $T = 0.01$}
  \end{subfigure}
  \begin{subfigure}[h]{0.3 \textwidth}
    \centering
    \includegraphics[height = 2.0 in]{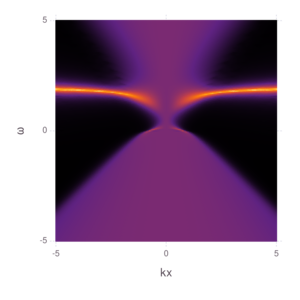}
    \caption{$\Phi = 0$, $\mu = -2$, $T = 0.01$}
  \end{subfigure}
  \begin{subfigure}[h]{0.3 \textwidth}
    \centering
    \includegraphics[height = 2.0 in]{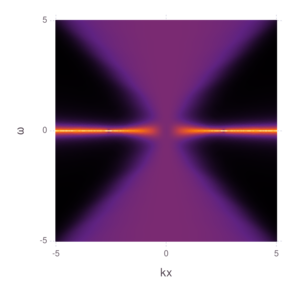}
    \caption{$\Phi = 0$, $\mu = 0$, $T = 0.2$}
  \end{subfigure}

  \begin{subfigure}[h]{0.3 \textwidth}
    \centering
    \includegraphics[height = 2.0 in]{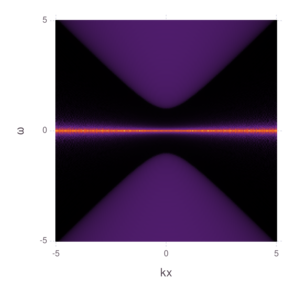}
    \caption{$\Phi = u$, $\mu = 0$, $T = 0.01$}
  \end{subfigure}
  \begin{subfigure}[h]{0.3 \textwidth}
    \centering
    \includegraphics[height = 2.0 in]{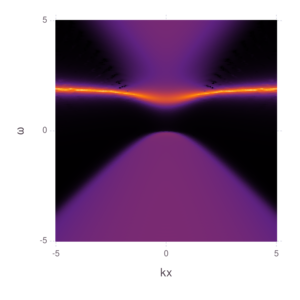}
    \caption{$\Phi = u$, $\mu = -2$, $T = 0.01$}
  \end{subfigure}
  \begin{subfigure}[h]{0.3 \textwidth}
    \centering
    \includegraphics[height = 2.0 in]{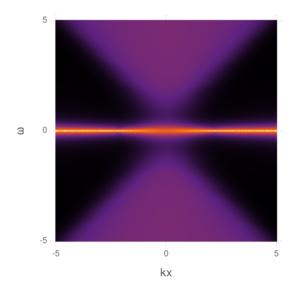}
    \caption{$\Phi = u$, $\mu = 0$, $T = 0.2$}
  \end{subfigure}

  \begin{subfigure}[h]{0.3 \textwidth}
    \centering
    \includegraphics[height = 2.0 in]{images/condensation_0.png}
    \caption{$\Phi = u^{2}$, $\mu = 0$, $T = 0.01$}
  \end{subfigure}
  \begin{subfigure}[h]{0.3 \textwidth}
    \centering
    \includegraphics[height = 2.0 in]{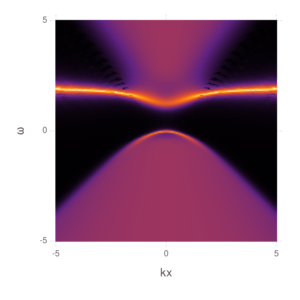}
    \caption{$\Phi = u^{2}$, $\mu = -2$, $T = 0.01$}
  \end{subfigure}
  \begin{subfigure}[h]{0.3 \textwidth}
    \centering
    \includegraphics[height = 2.0 in]{images/condensation_temp.png}
    \caption{$\Phi = u^{2}$, $\mu = 0$, $T = 0.2$}
  \end{subfigure}
   \caption{The effects of chemical potential  (b,e,h: the 2nd column)  and temperature (c,f,i: the 3rd column). 
  The first row figures (a,b,c) are for $\Phi=0$. The second  row (d,e,f) is for  the  scalar source, the
   third row (g,h,i)  are for the scalar condensation.    }
  \label{figure:TMeffect}
\end{figure}

\begin{figure}[t]
  \centering
  \begin{subfigure}[h]{0.6 \textwidth}
    \centering
    \includegraphics[height = 2.5 in]{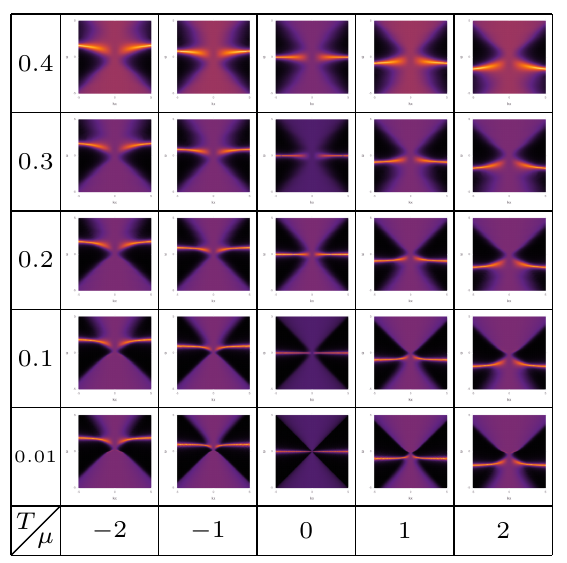}
    \caption{Non-Interacting ($\Phi = 0$)}   \label{figure:effects_of_chemical_potential_and_temperature_in_mixed_quantization_nonint}
  \end{subfigure}
  \begin{subfigure}[h]{0.45 \textwidth}
    \centering
    \includegraphics[height = 2.5 in]{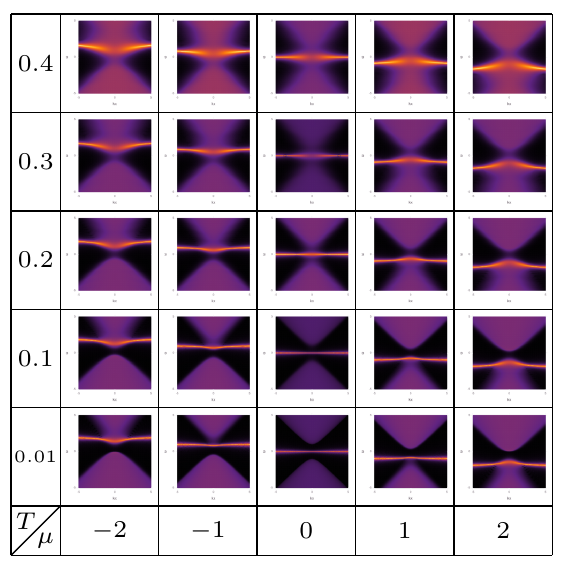}
    \caption{Source ($\Phi = \Phi_{0} u$ with $\Phi_{0} = 1$)}    \label{figure:effects_of_chemical_potential_and_temperature_in_mixed_quantization_source}
  \end{subfigure}
  \begin{subfigure}[h]{0.45 \textwidth}
    \centering
    \includegraphics[height = 2.5 in]{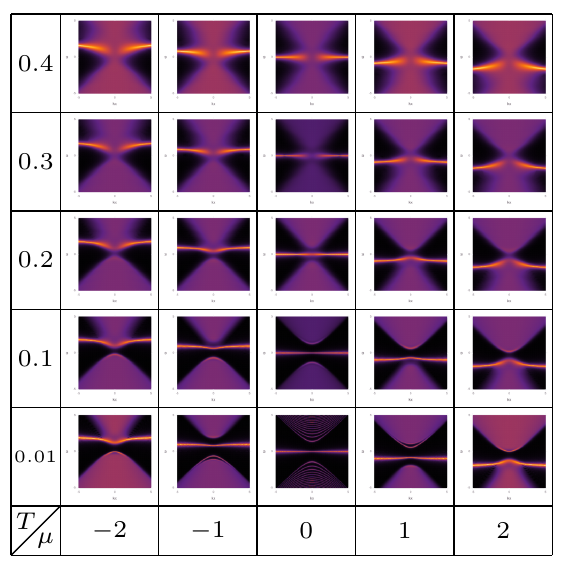}
    \caption{Condensation ($\Phi = \Phi_{0} u^{2}$ with $\Phi_{0} = 1$)}    \label{figure:effects_of_chemical_potential_and_temperature_in_mixed_quantization_condensation}
  \end{subfigure}
  \caption{
    Evolution of the effects of charge density and temperature in the mixed quantization.
    The tables show the spectral density functions for various density and temperature $T$ in $k_{x}$-$\omega$ plane
    (a) without Yukawa interaction,
    (b) with interaction of fermion with scalar source, and
    (c) with interaction of fermion with scalar condensation.
  }
  \label{figure:effects_of_chemical_potential_and_temperature_in_mixed_quantization}
\end{figure}

In this section, we study the effect of finite density and temperature numerically.
We use RN AdS black hole metric and gauge potential and probe scalar field as the background of the bulk fermion.
The charge density $\rho$ always comes with the chemical potential.
In the probe limit where we neglect the effect of the scalar on the background gravity and gauge fields, $\rho$ and $\mu$ are related by the regularity condition at the horizon $A_{u_{h}}=0$ where $A_{t}=\mu-\rho u=\mu(1-u/u_{h})$, so that $\rho=\mu/u_{h}$.
The fermion couplings to the metric and the gauge potential are still given by the previous setup.
One only needs to care eq. (\ref{DAt}).
The results are summarized in figure \ref{figure:TMeffect}.  

The effects of the chemical potential and the temperature can be summarized as follows.
Chemical potential itself just shifts the origin of the energy but it is followed by the density which makes the phenomena complicated.
The flat band is bent when it is touched by the Dirac bands.
For positive $\mu$, the flat band is bent as if it wants to avoid the Dirac spectra, and the spectral weight of the upper Dirac band is moved to the lower Dirac band as well as to the flat band.
See also figures \ref{figure:TMeffect} (b,e,h).
As a consequence, the lower Dirac band is strengthened  at the price of the  weakened lower one.
Such bending and the spectral weight transfer are the two most important strong correlation effects appearing in our investigation.

In the presence of the gap, the bending effect is very similar to the hybridization of the local f-band and parabolic band in the Kondo lattice system.
Indeed, what we see in figures \ref{figure:TMeffect} (b,e,h) is the holographic version of the hybridization effect, since it is due to the presence of the off-diagonal terms.

The effect of the temperature is to make the spectrum fuzzy \ref{figure:TMeffect} (c,f,i), which can be understood since the degrees of freedom will be spread due to the thermal excitations.
Therefore when the temperature $T$ is high enough, all the gaps will be closed by the population of fuzzy spectral weights.
It is truly interesting to see that the holographic calculation can reveal such effects at ease.
More evolutions of the spectral data as $\mu$ and $T$ vary are given in figure \ref{figure:effects_of_chemical_potential_and_temperature_in_mixed_quantization}.

\section{ Taking care of the gravity back reaction}
So far, all the calculations were first done in the probe limit. Here  we take care of the gravity back reaction to show that 
the result does not change the qualitative features of the probe limit calculations. 
It turns out that the calculation time for a one process takes more than 20 times of CPU time. Therefore, the probe limit calculation has certainly a merit at the discovery stage.

Our action is given by  
\begin{equation}
  S_{\text{background}} = \int \dd^{4} x \sqrt{-g} \left[ R + \frac{6}{L^{2}} - \frac{1}{4} F_{\mu \nu} F^{\mu \nu} + (\partial_{\mu} \Phi)^{2} + 2 \Phi^{2} \right],
\end{equation}
where $L = 1$.
When we take ansatz
\begin{equation}
  \dd s^{2} = \frac{1}{u^{2}} (-f \chi \dd t^{2} + \dd x^{2} + \dd y^{2}) + \frac{\dd u^{2}}{f u^{2}},
\end{equation}
\begin{equation}
  f = f(u), \quad
  \chi = \chi(u), \quad
  A = A_{t}(u) \dd t, \quad
  \Phi = \Phi(u),
\end{equation}
the equations of motion are as follows \cite{Hartnoll:2008kx}:
\begin{align}
  f' - \frac{3 f}{u} + \frac{3}{u} + \frac{f \chi'}{2 \chi} - \frac{A_{t}'^{2} u^{3}}{4 \chi} + \frac{\Phi^{2}}{u} = &\ 0, \\
  \chi' + \Phi'^{2} \chi u = &\ 0, \\
  A_{t}'' - \frac{\chi' A_{t}'}{2 \chi} = &\ 0, \\
  \Phi'' + \left( \frac{f'}{f} - \frac{2}{u} + \frac{\chi'}{2 \chi} \right) \Phi' + \frac{2}{f u^{2}} \Phi = &\ 0.
\end{align}
The asymptotic behavior of the background fields near the boundary is given by
\begin{align}
  A_{t} \approx &\ \mu - \rho u + \cdots, \\
  \Phi \approx &\ \Phi_{-} u + \Phi_{+} u^{2} + \cdots.
\end{align}

\begin{figure}[t]
  \centering

  \begin{subfigure}[h]{0.22 \textwidth}
    \centering
    \includegraphics[width = 1.5 in]{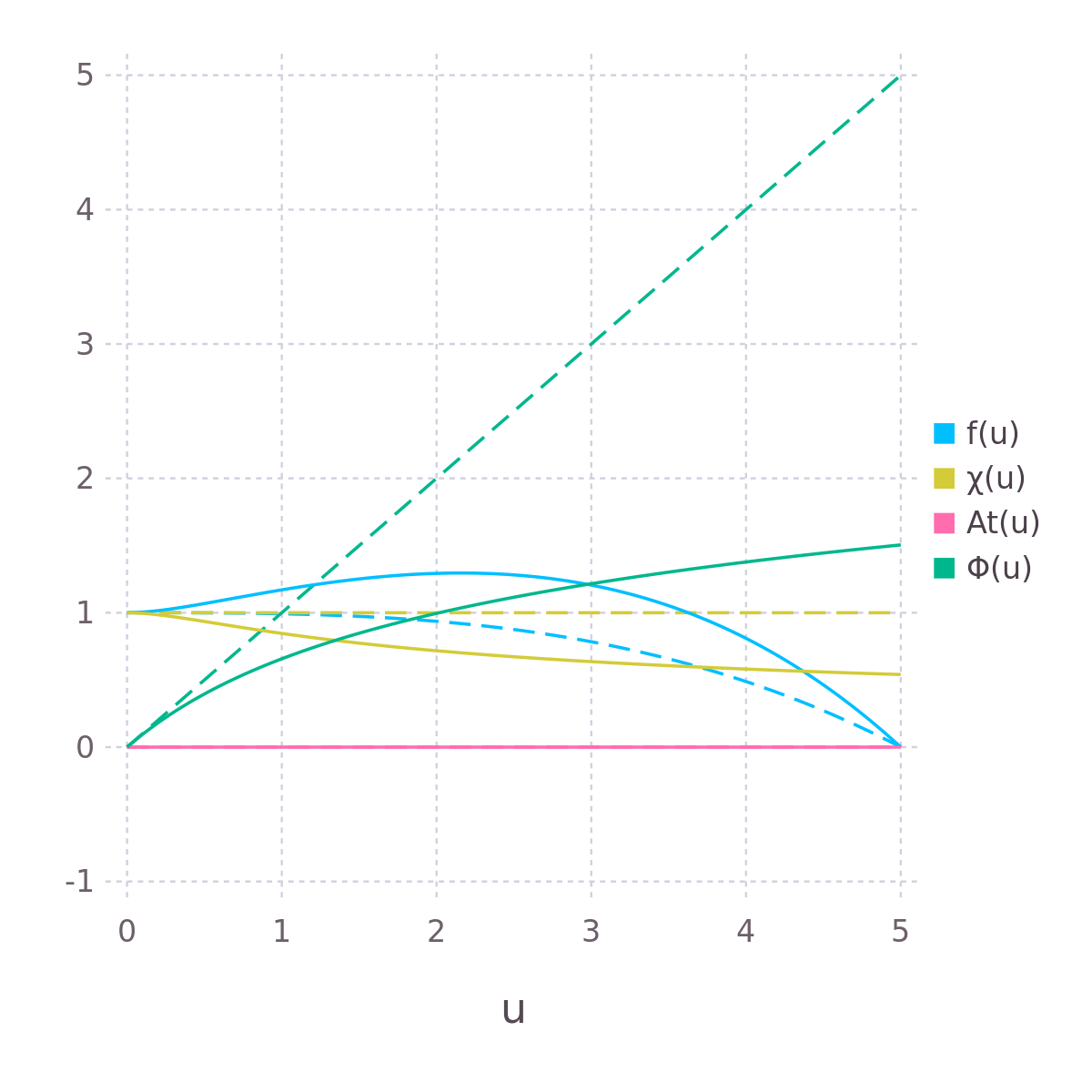}
    \caption{}
  \end{subfigure}
  \begin{subfigure}[h]{0.22 \textwidth}
    \centering
    \includegraphics[width = 1.5 in]{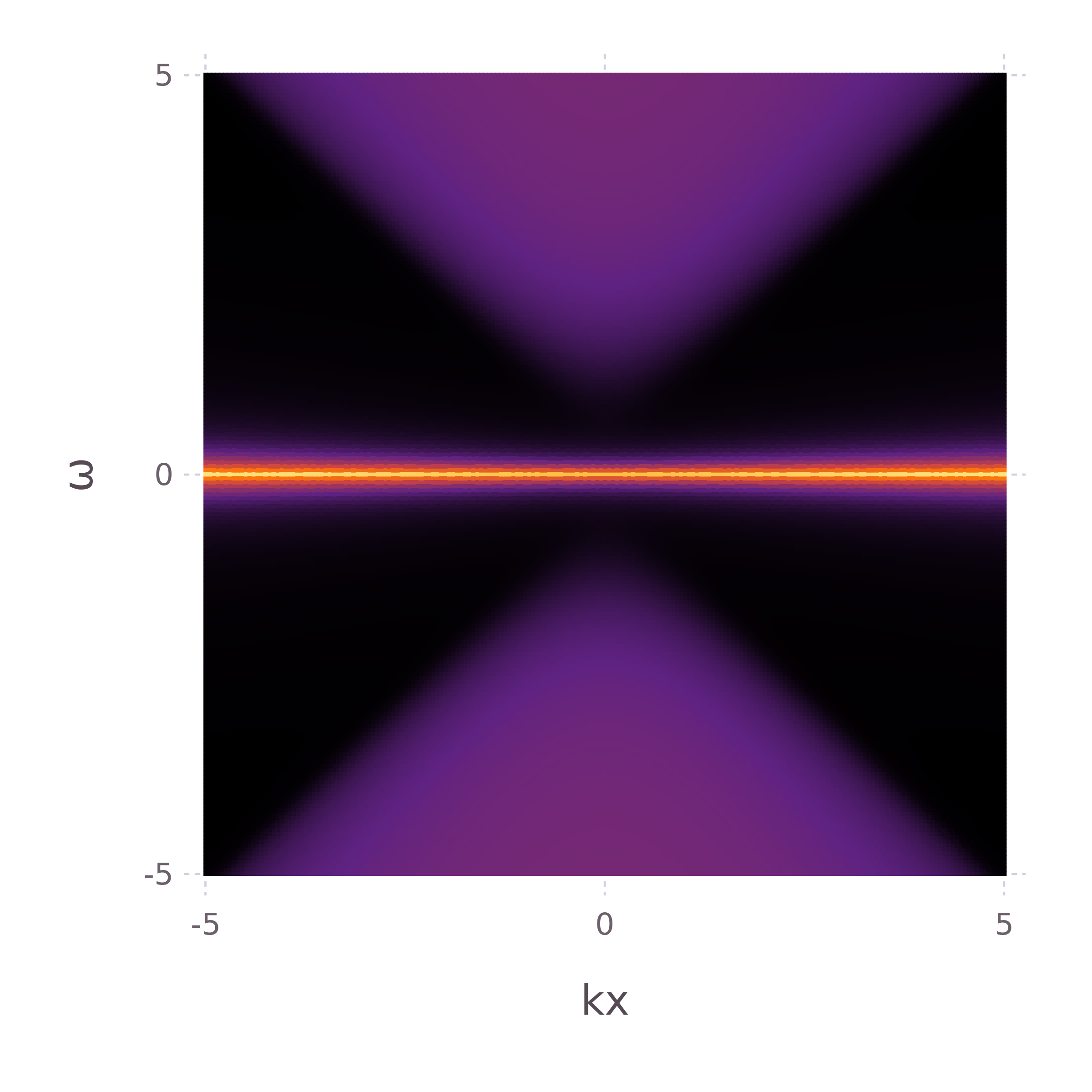}
    \caption{}
  \end{subfigure}
  \begin{subfigure}[h]{0.22 \textwidth}
    \centering
    \includegraphics[width = 1.5 in]{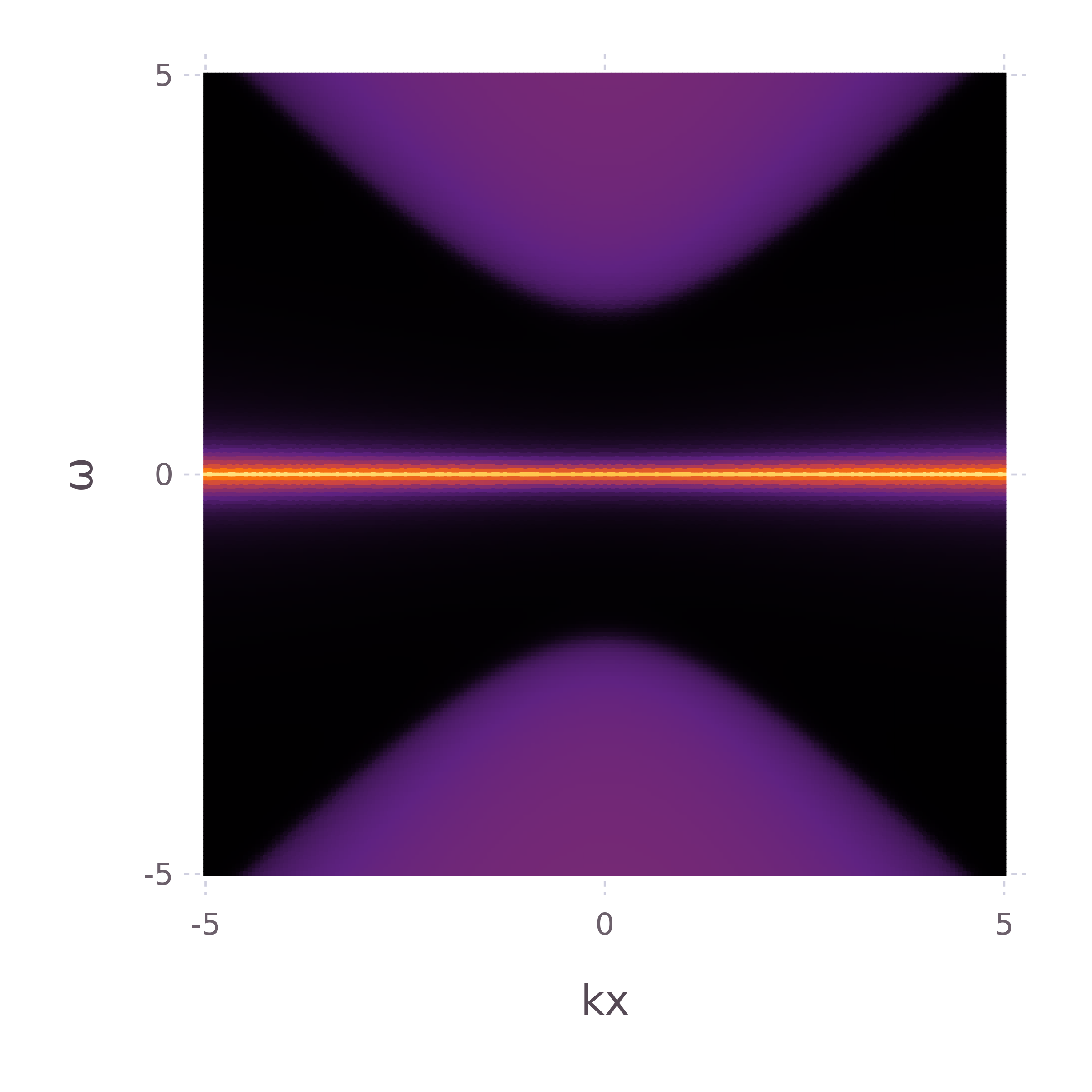}
    \caption{}
  \end{subfigure}
  \begin{subfigure}[h]{0.22 \textwidth}
    \centering
    \includegraphics[width = 1.5 in]{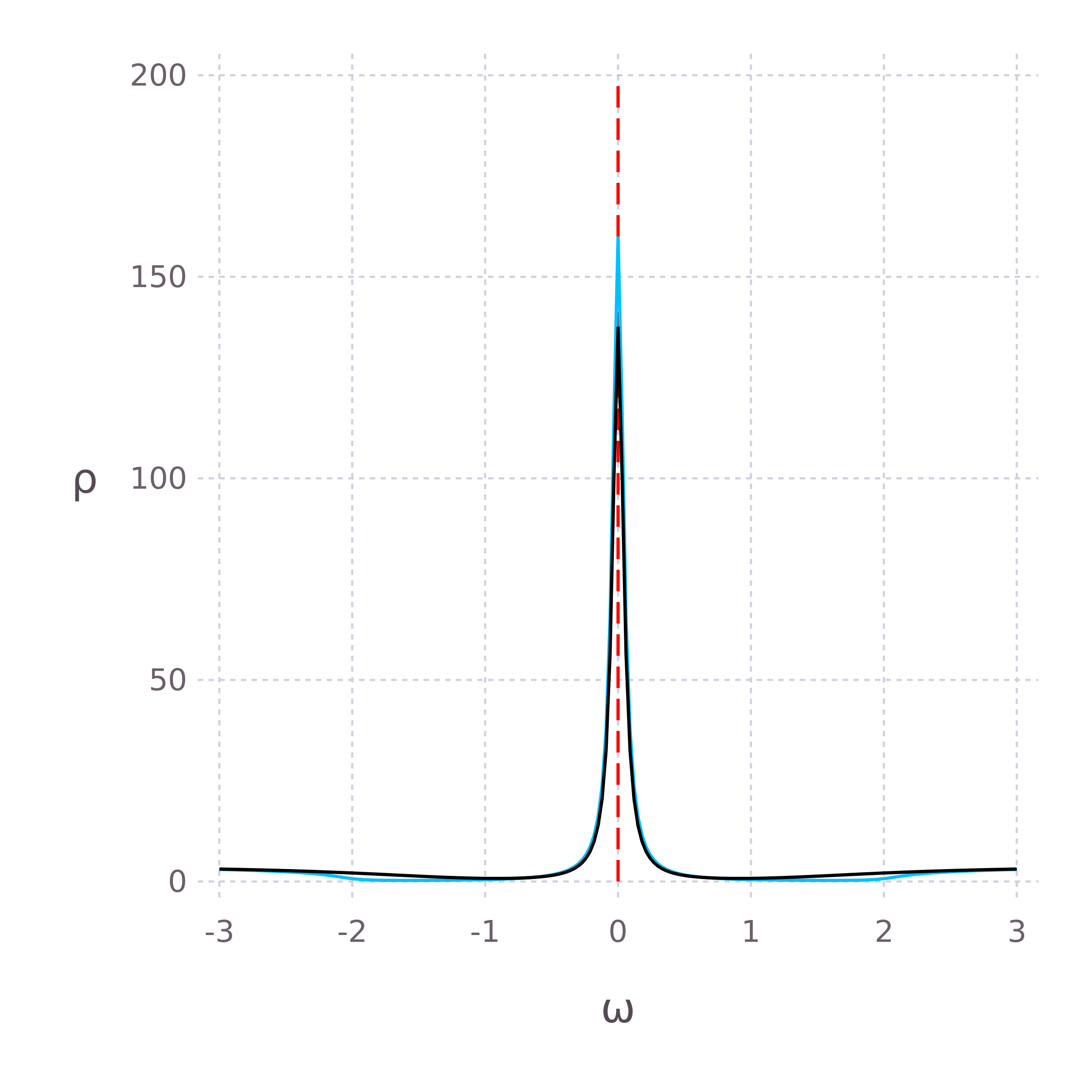}
    \caption{}
  \end{subfigure}
  \\
  \begin{subfigure}[h]{0.22 \textwidth}
    \centering
    \includegraphics[width = 1.5 in]{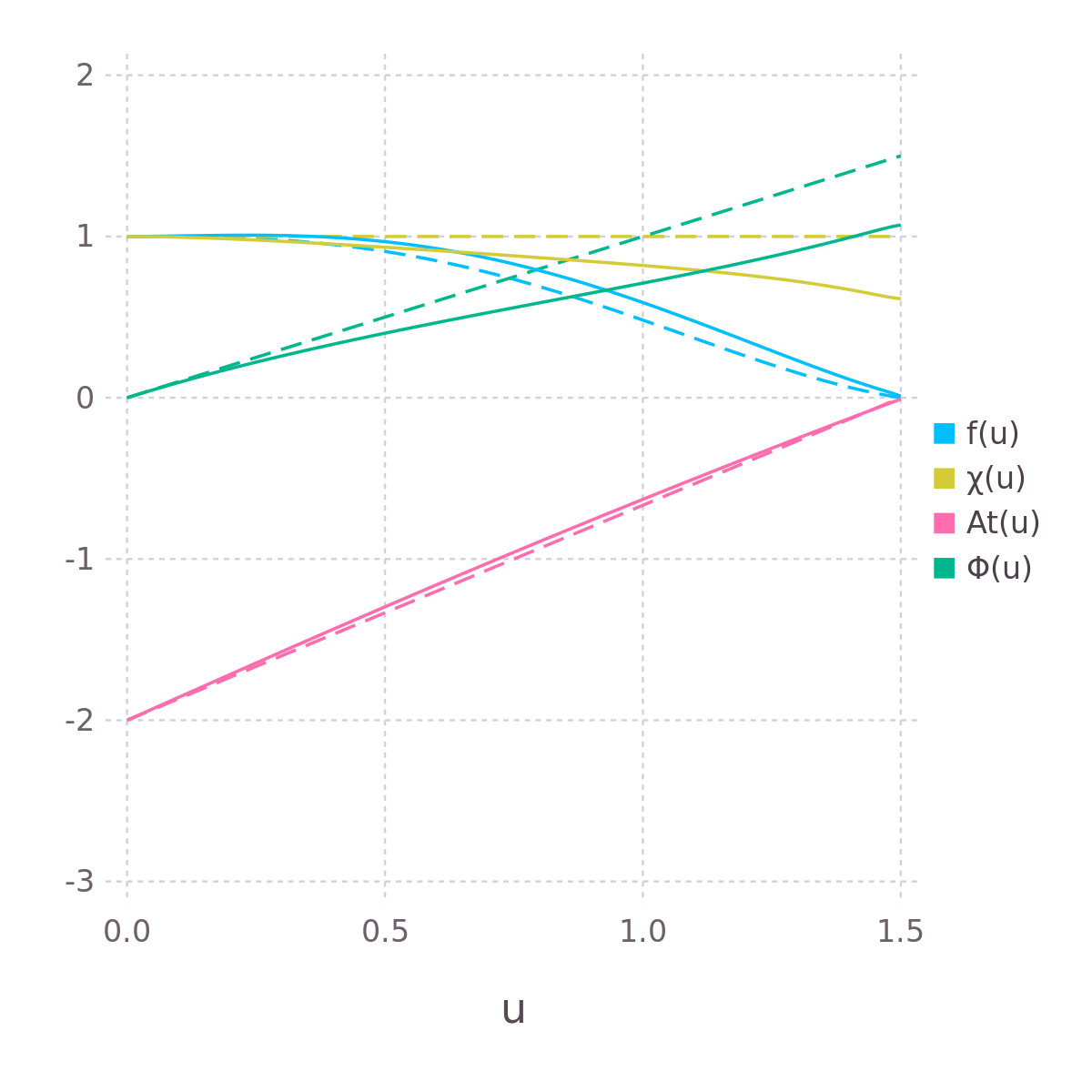}
    \caption{}
  \end{subfigure}
  \begin{subfigure}[h]{0.22 \textwidth}
    \centering
    \includegraphics[width = 1.5 in]{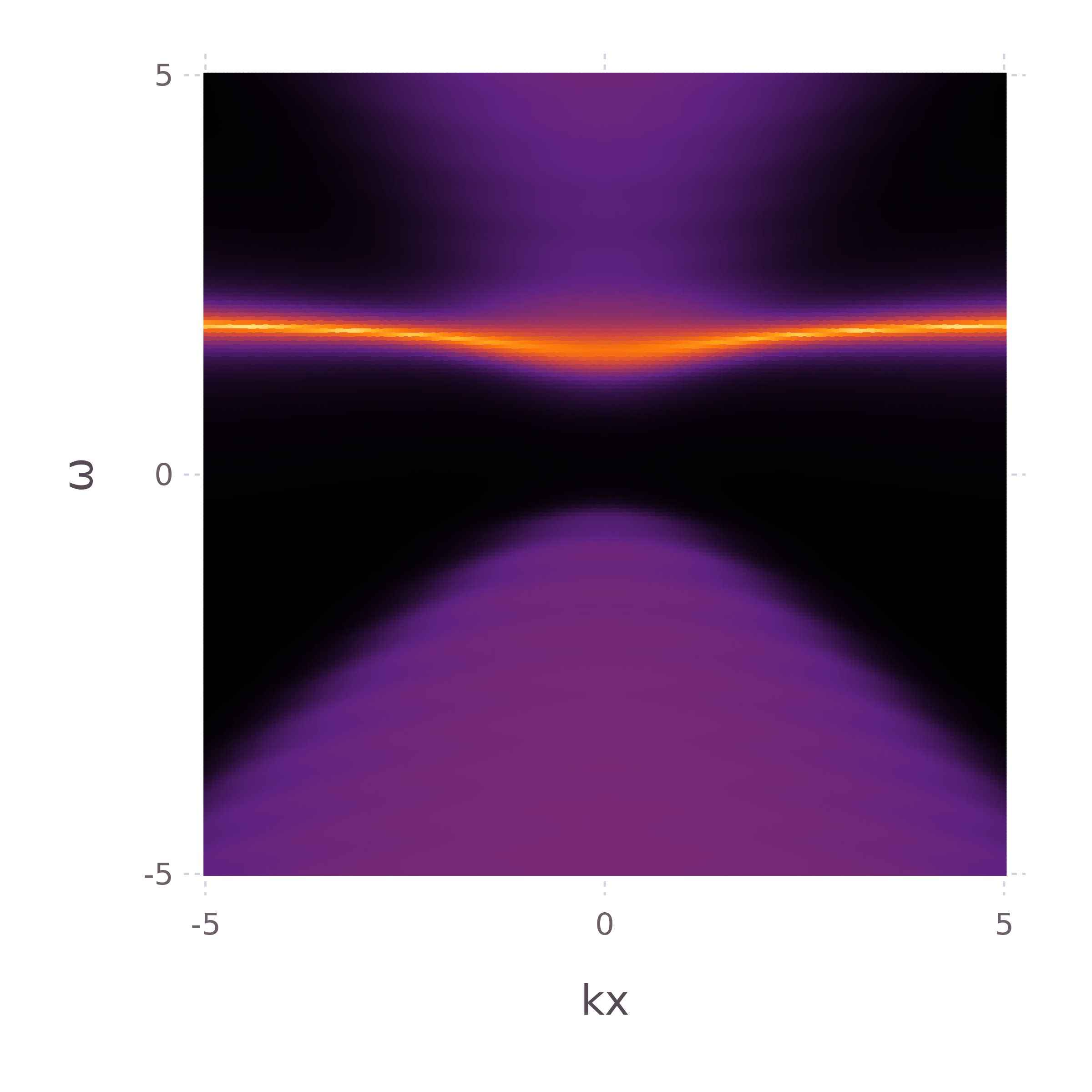}
    \caption{}
  \end{subfigure}
  \begin{subfigure}[h]{0.22 \textwidth}
    \centering
    \includegraphics[height = 1.5 in]{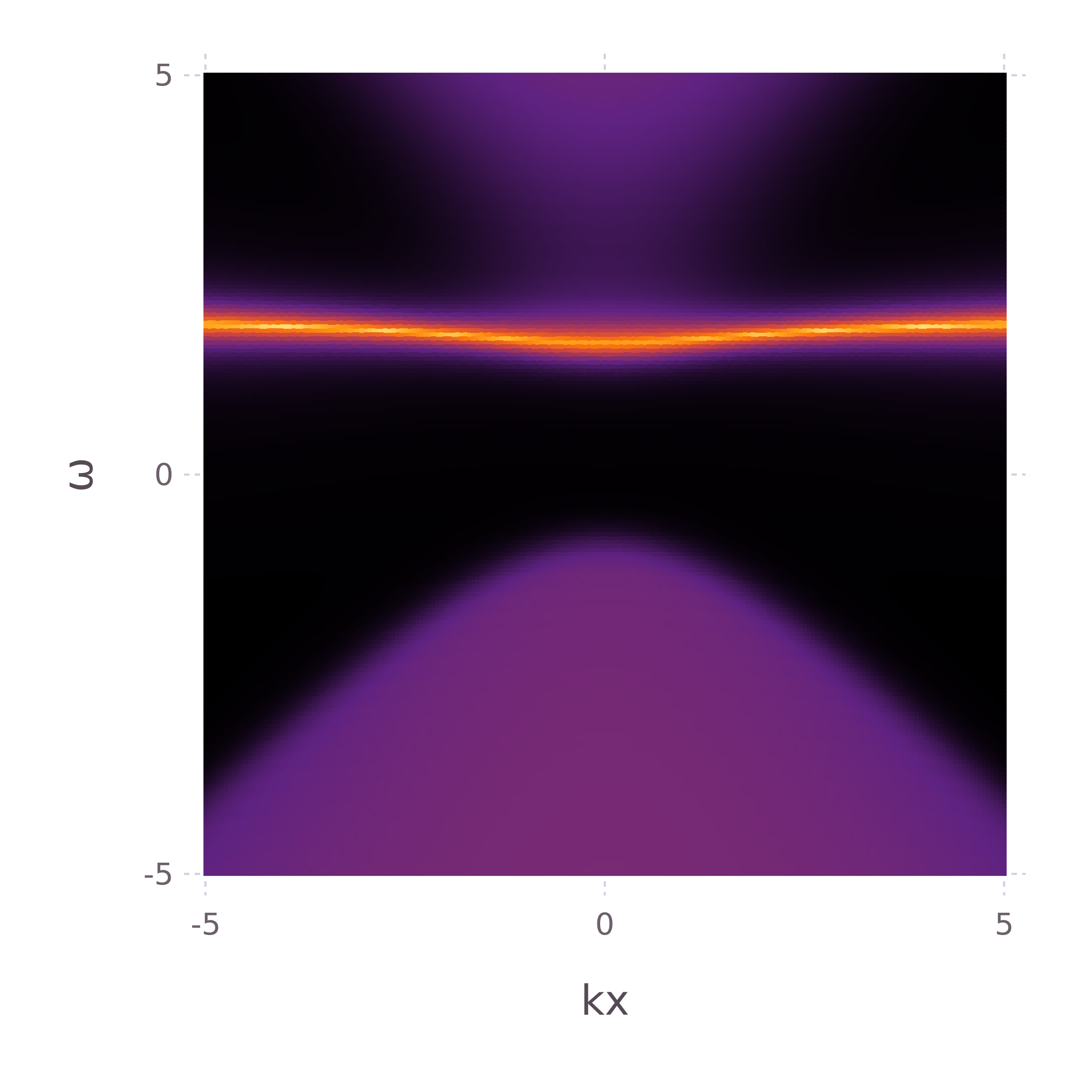}
    \caption{}
  \end{subfigure}
  \begin{subfigure}[h]{0.22 \textwidth}
    \centering
    \includegraphics[height = 1.5 in]{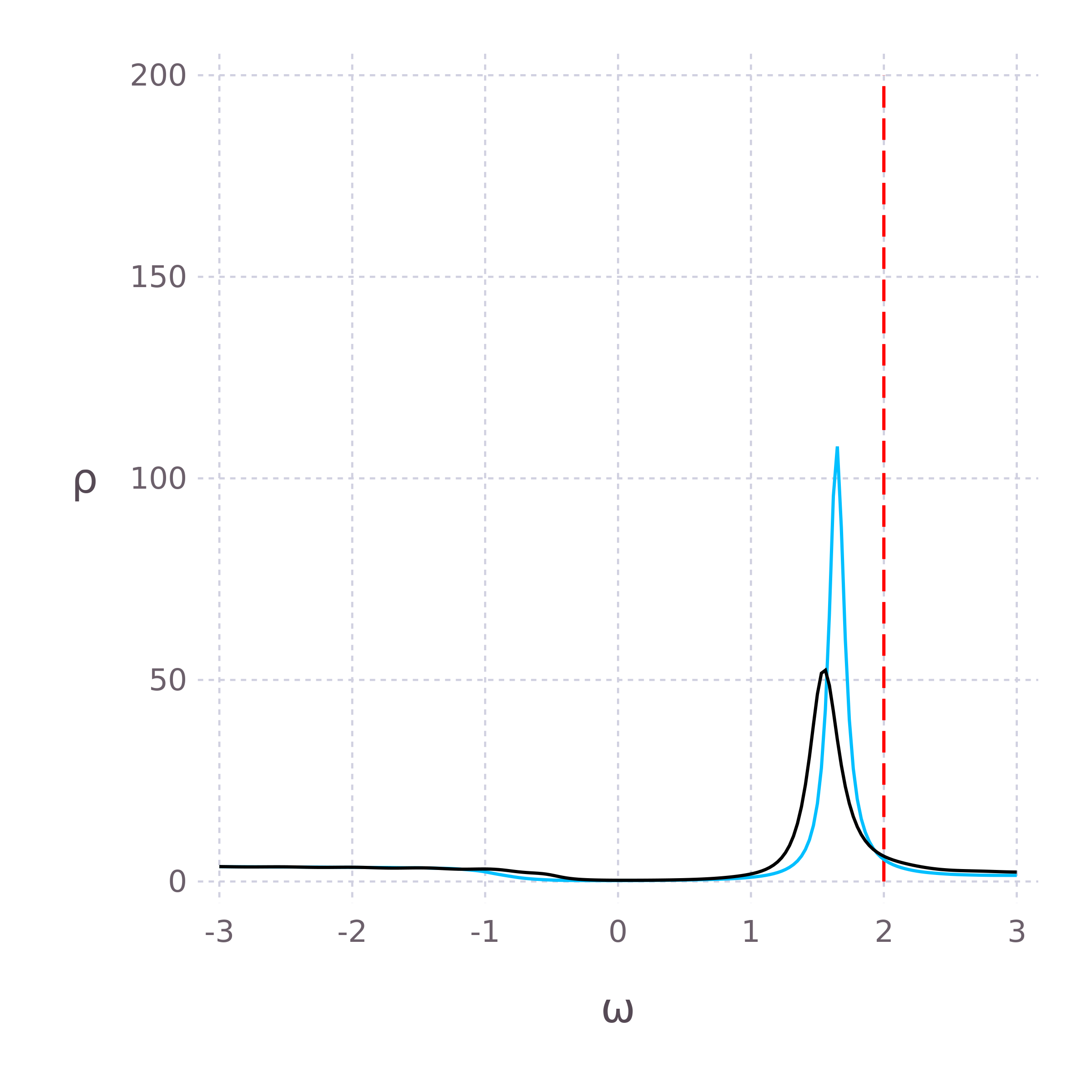}
    \caption{}
  \end{subfigure}
  \\
  \begin{subfigure}[h]{0.22 \textwidth}
    \centering
    \includegraphics[width = 1.5 in]{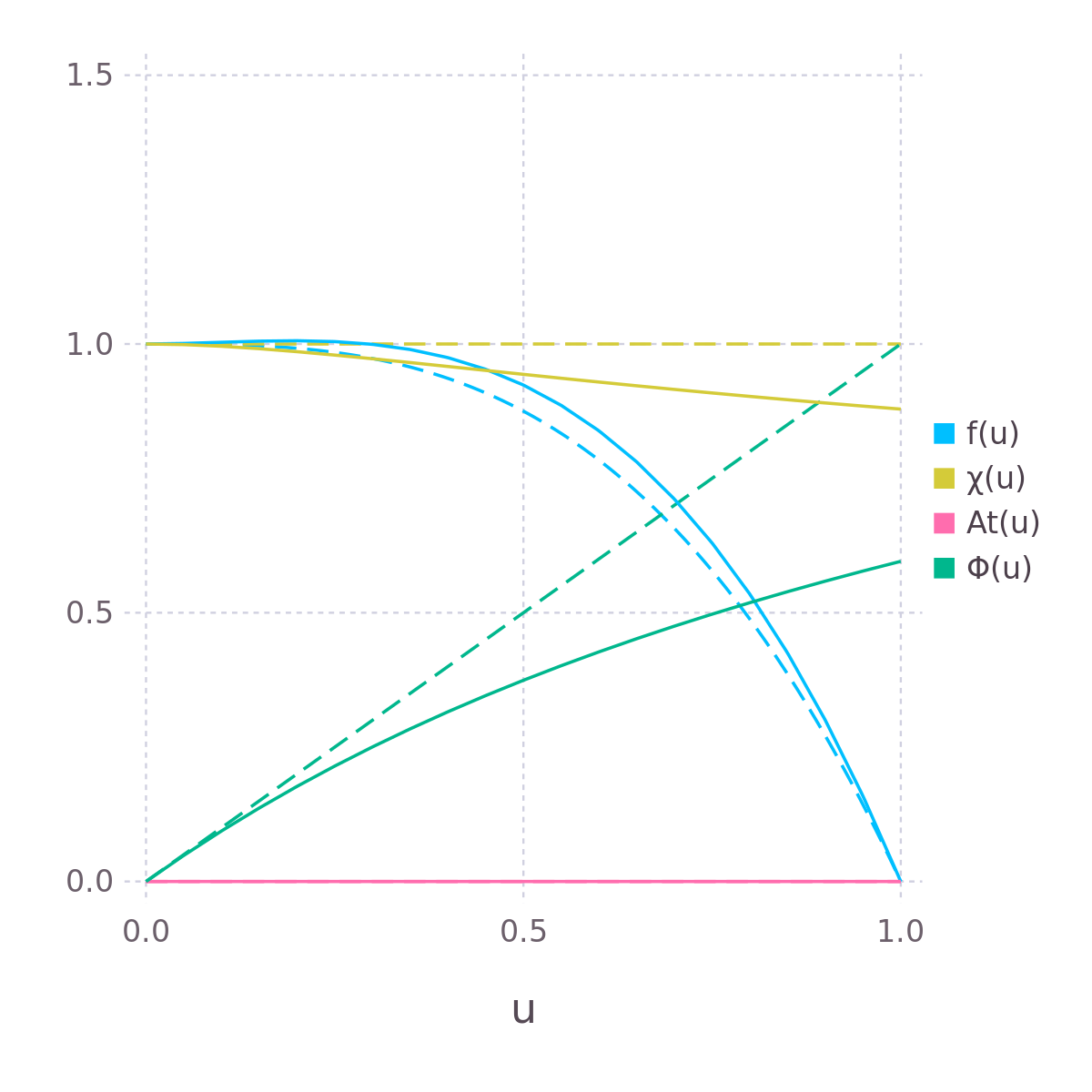}
    \caption{}
  \end{subfigure}
  \begin{subfigure}[h]{0.22 \textwidth}
    \centering
    \includegraphics[width = 1.5 in]{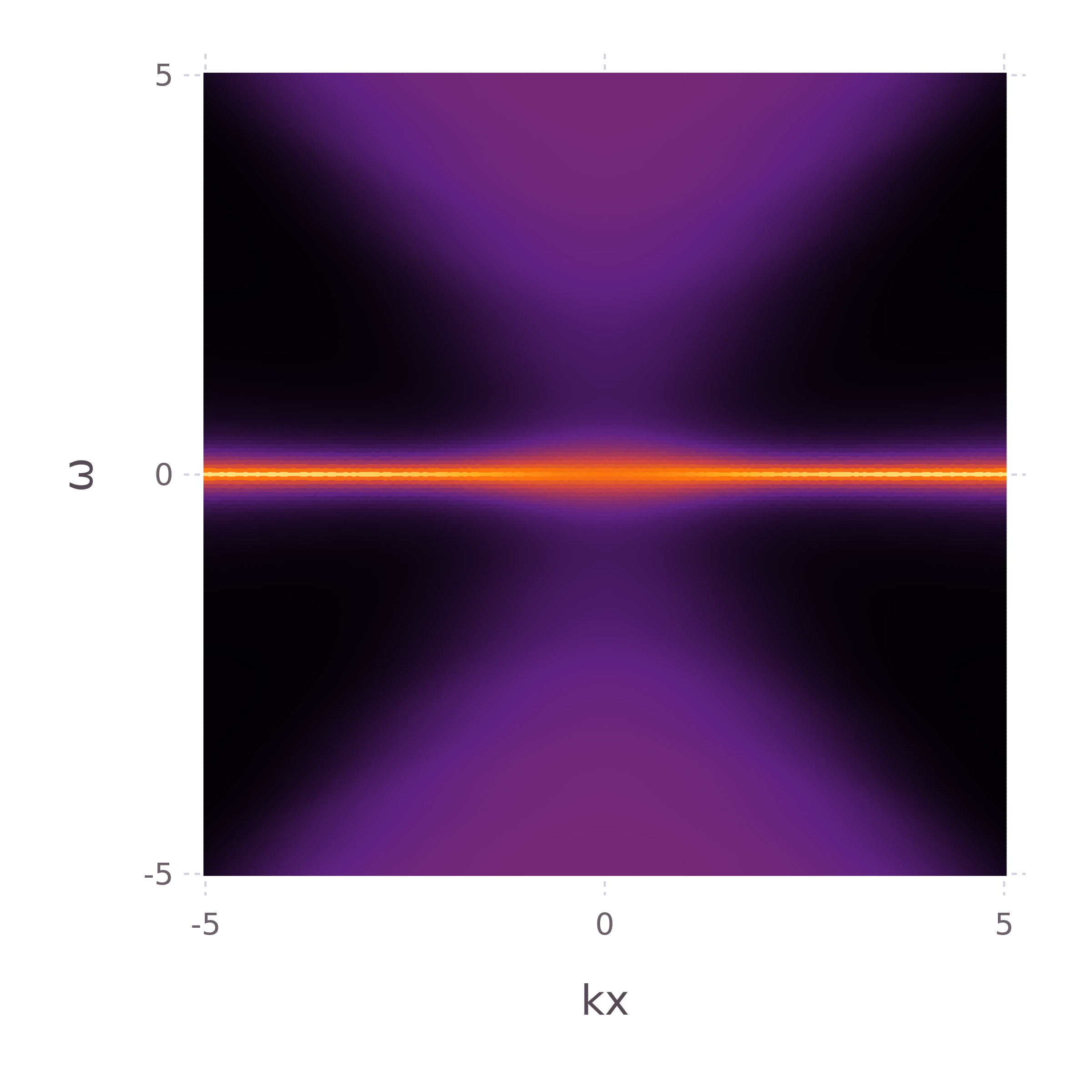}
    \caption{}
  \end{subfigure}
  \begin{subfigure}[h]{0.22 \textwidth}
    \centering
    \includegraphics[height = 1.5 in]{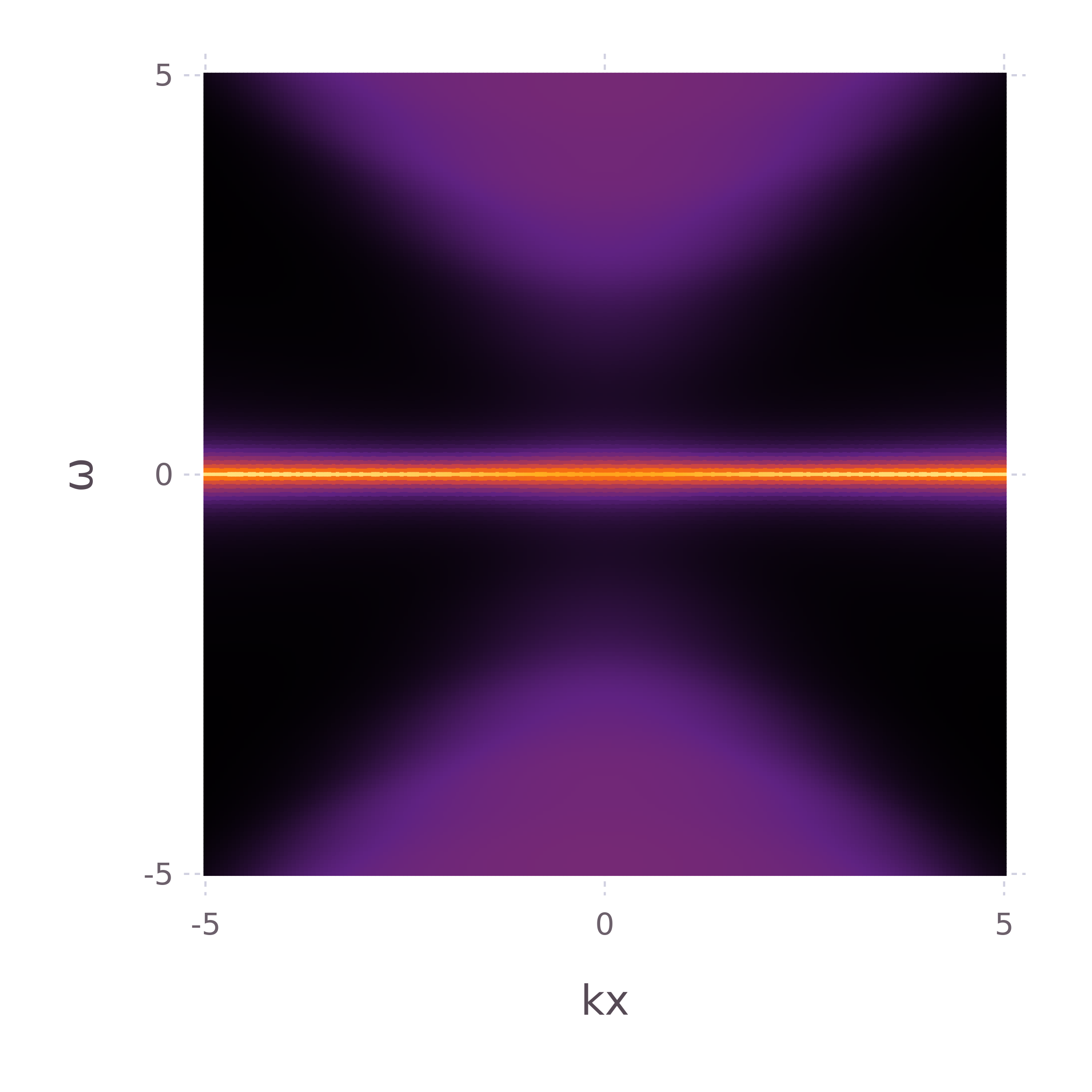}
    \caption{}
  \end{subfigure}
  \begin{subfigure}[h]{0.22 \textwidth}
    \centering
    \includegraphics[height = 1.5 in]{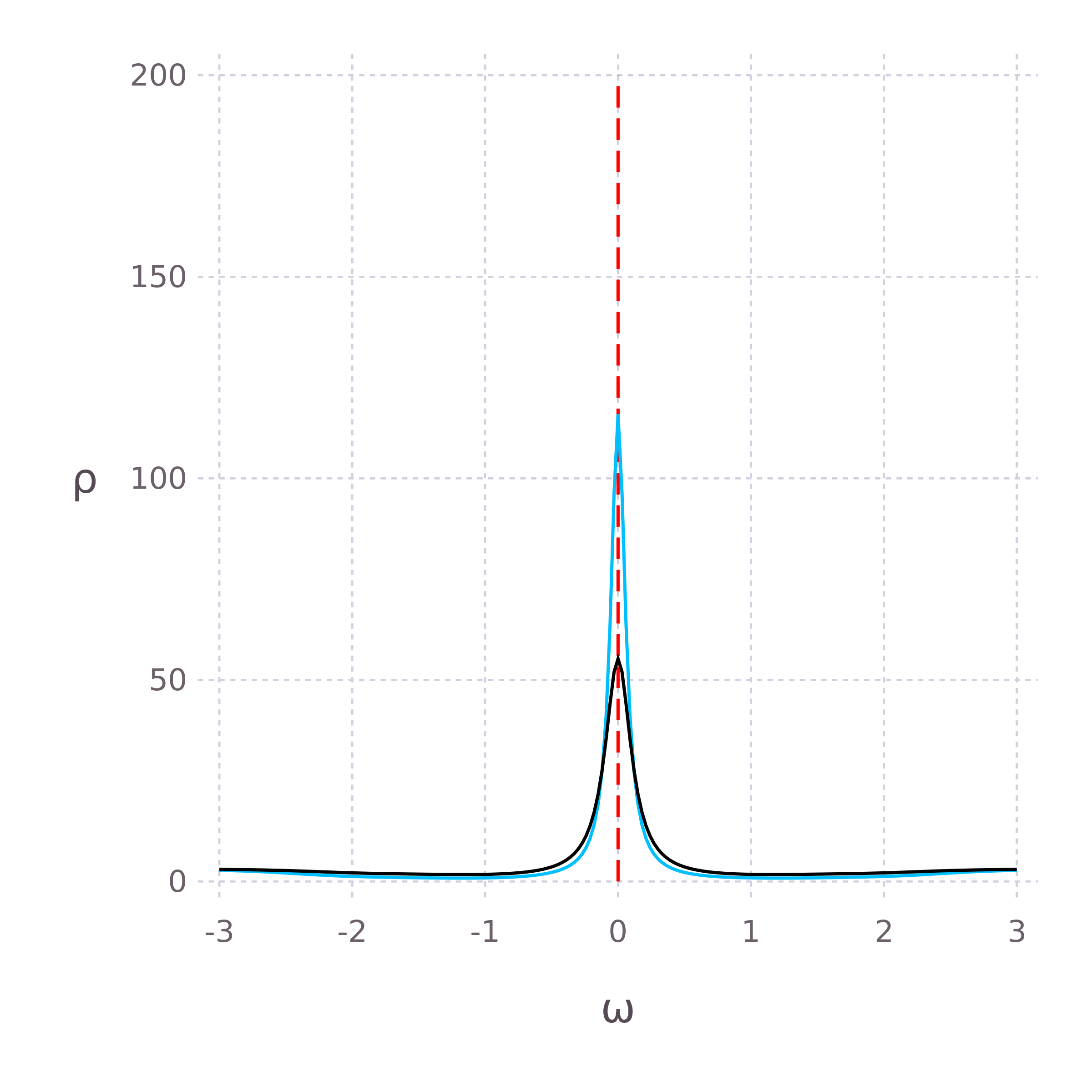}
    \caption{}
  \end{subfigure}

  \caption{Comparison of calculations with and without back reaction.
    (a,e,i) Background fields with (solid line) and without (dashed line) back reaction.
    (b,f,j) Fermion spectra with back reaction.
    (c,g,k) Fermion spectra without back reaction.
    (d,h,l) Spectral function $\rho$ on $k_{x} = k_{y} = 0$ with (black line) and without (blue line) back reaction, and position of undeformed flat band (red dashed line).
    We set parameters as $m = 0$, $q = 1$, $\lambda = 2$ (all); $\mu = 0$, $u_{h} = 5$ (a,b,c,d); $\mu = -2$, $u_{h} = 1.5$ (e,f,g,h); $\mu = 0$, $u_{h} = 1$ (i,j,k,l).
  }
  \label{figure:comparison_of_calculations_with_and_without_back_reaction}
\end{figure}

In principle, we should impose boundary condition at $u = 0$ and $u = u_{h}$ as follows:
\begin{align}
  f(u_{h}) = &\ 0, &
  \chi(0) = &\ 1, \\
  \mu = &\ \text{control parameter}, &
  A(u_{h}) = &\ 0, \\
  \Phi_{-} = &\ \text{control parameter}, &
  \Phi(u_{h}) = &\ \text{finite}.
\end{align}
However, to avoid infinity in numerical calculation, we cut off domain as $u \in [\epsilon, u_{h} - \epsilon]$ with some small value $\epsilon$, and then set boundary condition as follows:
\begin{align}
  f(u_{h} - \epsilon) + \epsilon f'(u_{h} - \epsilon) = &\ 0, &
  \chi(\epsilon) - \epsilon \chi'(\epsilon) - 1 = &\ 0, \\
  A_{t}(\epsilon) - \epsilon A_{t}'(\epsilon) - \mu = &\ 0, &
  A_{t}(u_{h} - \epsilon) + \epsilon A_{t}'(u_{h} - \epsilon) = &\ 0, \\
  \Phi(\epsilon) - \frac{1}{2} \epsilon \Phi'(\epsilon) - \frac{1}{2} \Phi_{-} \epsilon = &\ 0, &
  \epsilon^{2} \Phi''(u_{h} - \epsilon) = &\ 0.
\end{align}
We use the mono-implicit Runge-Kutta (MIRK) method with initial guess
\begin{equation}
  f^{(i)} = 1 - \left( \frac{u}{u_{h}} \right)^{3}, \quad
  \chi^{(i)} = 1, \quad
  A_{t}^{(i)} = \mu \left( 1 - \frac{u}{u_{h}} \right), \quad
  \Phi^{(i)} = 0.
\end{equation}
The action of fermion perturbation $\psi$ is given by
\begin{equation}
  S_{\psi} = \int \dd^{4} x \sqrt{-g} i \bar{\psi} \left[ \frac{1}{2} \left( \overrightarrow{\slashed{D}} - \overleftarrow{\slashed{D}} \right) - m - i \lambda (i \mathbb{I}_{4} \Phi) \right] \psi + \frac{1}{2} \int \dd^{3} x \sqrt{-h} \bar{\psi} \Gamma^{\underbar{x} \underbar{y}} \psi.
\end{equation}
Figure \ref{figure:comparison_of_calculations_with_and_without_back_reaction} shows calculation with back reaction ($\Phi_{-} = 1$), and approximated calculation without back reation [$f = 1 - (1 + u_{h}^{2} \mu^{2} / 4) (u / u_{h})^{3} + (u_{h}^{2} \mu^{2} / 4) (u / u_{h})^{4}$, $\chi = 1$, $A_{t} = \mu (1 - u / u_{h})$, and $\Phi = u$].

\section{Discussion}
In this paper, we realized the flat band separated from the other band by gapping the Dirac band from the band in the band structure of the Lieb lattice.  Such an isolated flat band will be able to play the role of the hydrogen atom of strongly correlated systems, the simplest system where various interesting phenomena caused by the strong correlation can be observed.

We identify the role of the boundary condition used in    Laia-Tong model.
It play the role of creating a compact closed state  by taking the projection to two degrees of freedom whose contribution to kinetic energy are cancelled by each other.  That is, the mixed BC   corresponds to creating a compact localized state \cite{maimaiti2017compact} in the lattice that creates the flat band.
So our interpretation is that, before we impose the BC, only the Dirac band exists.
With the BC, a flat band is created, and the interaction causes the spectral transfer so that most of the spectral weight of the Dirac band is transferred to the flat band, which is represented by the holographic spectral function in the mixed quantization.

We also explained the appearance of many bands in the holographic theories
 and  studied the density effect. 
All the calculations were first done in the probe limit and later justify that 
the gravity back reaction does not change the qualitative features. 

 Another point to make here is that the tight-binding model of the Lieb lattice is known to be unstable in any generic perturbation so that there is no stand-alone Lieb lattice realized in real materials, while the holographic system is a stable system.
So we should identify the Laia-Tong model as a stabilized and realizable model rather than the unstable tight-binding model.  
It would be interesting to realize other lattice models in holographic setups.
However, more interesting problems will be to calculate the physical observables other than the band structure itself, like the conductivity and quantum metric in the presence of the flat band.
It is also interesting to consider the parabolically touching band with the flat band and its holographic realization.
It would also be interesting to study   the holographic system where both the Dirac band and the flat bands are of particle spectrum since, here,  the Dirac band is an unparticle spectrum.
We will return to these problems in future works.

\vspace{ 0.2cm}
\acknowledgments
This  work is supported by Mid-career Researcher Program through the National Research Foundation of Korea grant No. NRF-2021R1A2B5B02002603. 
We also  thank the APCTP for the hospitality during the focus program, “Quantum Matter and   Entanglement with Holography”, where part of this work was discussed.

\bibliographystyle{jhep}
\bibliography{lieb_refs.bib}

\providecommand{\href}[2]{#2}\begingroup\raggedright\begin{thebibliography}{10}

\bibitem{Cao:2018:10.1038_nature26154}
Y.~Cao, V.~Fatemi, A.~Demir, S.~Fang, S.L.~Tomarken, J.Y.~Luo et~al.,
  \emph{Correlated insulator behaviour at half-filling in magic-angle graphene
  superlattices}, \href{https://doi.org/10.1038/nature26154}{\emph{Nature}
  {\bfseries 556} (2018) 80}.

\bibitem{Cao:2018:10.1038_nature26160}
Y.~Cao, V.~Fatemi, S.~Fang, K.~Watanabe, T.~Taniguchi, E.~Kaxiras et~al.,
  \emph{Unconventional superconductivity in magic-angle graphene
  superlattices}, \href{https://doi.org/10.1038/nature26160}{\emph{Nature}
  {\bfseries 556} (2018) 43}.

\bibitem{Mielke:1991:10.1088_0305-4470_24_14_018}
A.~Mielke, \emph{Ferromagnetism in the hubbard model on line graphs and further
  considerations},
  \href{https://doi.org/10.1088/0305-4470/24/14/018}{\emph{Journal of Physics
  A: Mathematical and General} {\bfseries 24} (1991) 3311}.

\bibitem{Mielke:1992:10.1088_0305-4470_25_16_011}
A.~Mielke, \emph{Exact ground states for the hubbard model on the kagome
  lattice}, \href{https://doi.org/10.1088/0305-4470/25/16/011}{\emph{Journal of
  Physics A: Mathematical and General} {\bfseries 25} (1992) 4335}.

\bibitem{Tasaki:1992:10.1103_PhysRevLett.69.1608}
H.~Tasaki, \emph{Ferromagnetism in the hubbard models with degenerate
  single-electron ground states},
  \href{https://doi.org/10.1103/PhysRevLett.69.1608}{\emph{Phys. Rev. Lett.}
  {\bfseries 69} (1992) 1608}.

\bibitem{Lieb:1989:10.1103_PhysRevLett.62.1201}
E.H.~Lieb, \emph{Two theorems on the hubbard model},
  \href{https://doi.org/10.1103/PhysRevLett.62.1201}{\emph{Phys. Rev. Lett.}
  {\bfseries 62} (1989) 1201}.

\bibitem{Costa:2016:10.1103_PhysRevB.94.155107}
N.C.~Costa, T.~Mendes-Santos, T.~Paiva, R.R.d.~Santos and R.T.~Scalettar,
  \emph{Ferromagnetism beyond lieb's theorem},
  \href{https://doi.org/10.1103/PhysRevB.94.155107}{\emph{Phys. Rev. B}
  {\bfseries 94} (2016) 155107}.

\bibitem{Tamura:2002:10.1103_PhysRevB.65.085324}
H.~Tamura, K.~Shiraishi, T.~Kimura and H.~Takayanagi, \emph{Flat-band
  ferromagnetism in quantum dot superlattices},
  \href{https://doi.org/10.1103/PhysRevB.65.085324}{\emph{Phys. Rev. B}
  {\bfseries 65} (2002) 085324}.

\bibitem{Imada:2000:10.1103_PhysRevLett.84.143}
M.~Imada and M.~Kohno, \emph{Superconductivity from flat dispersion designed in
  doped mott insulators},
  \href{https://doi.org/10.1103/PhysRevLett.84.143}{\emph{Phys. Rev. Lett.}
  {\bfseries 84} (2000) 143}.

\bibitem{Kopnin:2011:10.1103_PhysRevB.83.220503}
N.B.~Kopnin, T.T.~Heikkil\"a and G.E.~Volovik, \emph{High-temperature surface
  superconductivity in topological flat-band systems},
  \href{https://doi.org/10.1103/PhysRevB.83.220503}{\emph{Phys. Rev. B}
  {\bfseries 83} (2011) 220503}.

\bibitem{Julku:2016:10.1103_PhysRevLett.117.045303}
A.~Julku, S.~Peotta, T.I.~Vanhala, D.-H.~Kim and P.~T\"orm\"a, \emph{Geometric
  origin of superfluidity in the lieb-lattice flat band},
  \href{https://doi.org/10.1103/PhysRevLett.117.045303}{\emph{Phys. Rev. Lett.}
  {\bfseries 117} (2016) 045303}.

\bibitem{Dauphin:2016:10.1103_PhysRevA.93.043611}
A.~Dauphin, M.~M\"uller and M.A.~Martin-Delgado, \emph{Quantum simulation of a
  topological mott insulator with rydberg atoms in a lieb lattice},
  \href{https://doi.org/10.1103/PhysRevA.93.043611}{\emph{Phys. Rev. A}
  {\bfseries 93} (2016) 043611}.

\bibitem{Katsura:2010:10.1209_0295-5075_91_57007}
H.~Katsura, I.~Maruyama, A.~Tanaka and H.~Tasaki, \emph{Ferromagnetism in the
  hubbard model with topological/non-topological flat bands},
  \href{https://doi.org/10.1209/0295-5075/91/57007}{\emph{{EPL} (Europhysics
  Letters)} {\bfseries 91} (2010) 57007}.

\bibitem{Green:2010:10.1103_PhysRevB.82.075104}
D.~Green, L.~Santos and C.~Chamon, \emph{Isolated flat bands and spin-1 conical
  bands in two-dimensional lattices},
  \href{https://doi.org/10.1103/PhysRevB.82.075104}{\emph{Phys. Rev. B}
  {\bfseries 82} (2010) 075104}.

\bibitem{Tang:2011:10.1103_PhysRevLett.106.236802}
E.~Tang, J.-W.~Mei and X.-G.~Wen, \emph{High-temperature fractional quantum
  hall states},
  \href{https://doi.org/10.1103/PhysRevLett.106.236802}{\emph{Phys. Rev. Lett.}
  {\bfseries 106} (2011) 236802}.

\bibitem{Sun:2011:10.1103_PhysRevLett.106.236803}
K.~Sun, Z.~Gu, H.~Katsura and S.~Das~Sarma, \emph{Nearly flatbands with
  nontrivial topology},
  \href{https://doi.org/10.1103/PhysRevLett.106.236803}{\emph{Phys. Rev. Lett.}
  {\bfseries 106} (2011) 236803}.

\bibitem{Neupert:2011:10.1103_PhysRevLett.106.236804}
T.~Neupert, L.~Santos, C.~Chamon and C.~Mudry, \emph{Fractional quantum hall
  states at zero magnetic field},
  \href{https://doi.org/10.1103/PhysRevLett.106.236804}{\emph{Phys. Rev. Lett.}
  {\bfseries 106} (2011) 236804}.

\bibitem{Wang:2011:10.1103_PhysRevLett.107.146803}
Y.-F.~Wang, Z.-C.~Gu, C.-D.~Gong and D.N.~Sheng, \emph{Fractional quantum hall
  effect of hard-core bosons in topological flat bands},
  \href{https://doi.org/10.1103/PhysRevLett.107.146803}{\emph{Phys. Rev. Lett.}
  {\bfseries 107} (2011) 146803}.

\bibitem{Sheng:2011:10.1038_ncomms1380}
D.~Sheng, Z.-C.~Gu, K.~Sun and L.~Sheng, \emph{Fractional quantum hall effect
  in the absence of landau levels},
  \href{https://doi.org/10.1038/ncomms1380}{\emph{Nature Communications}
  {\bfseries 2} (2011) }.

\bibitem{Weeks:2010:10.1103_PhysRevB.82.085310}
C.~Weeks and M.~Franz, \emph{Topological insulators on the lieb and perovskite
  lattices}, \href{https://doi.org/10.1103/PhysRevB.82.085310}{\emph{Phys. Rev.
  B} {\bfseries 82} (2010) 085310}.

\bibitem{Guo:2009:10.1103_PhysRevB.80.113102}
H.-M.~Guo and M.~Franz, \emph{Topological insulator on the kagome lattice},
  \href{https://doi.org/10.1103/PhysRevB.80.113102}{\emph{Phys. Rev. B}
  {\bfseries 80} (2009) 113102}.

\bibitem{Goldman:2011:10.1103_PhysRevA.83.063601}
N.~Goldman, D.F.~Urban and D.~Bercioux, \emph{Topological phases for fermionic
  cold atoms on the lieb lattice},
  \href{https://doi.org/10.1103/PhysRevA.83.063601}{\emph{Phys. Rev. A}
  {\bfseries 83} (2011) 063601}.

\bibitem{Beugeling:2012:10.1103_PhysRevB.86.195129}
W.~Beugeling, J.C.~Everts and C.~Morais~Smith, \emph{Topological phase
  transitions driven by next-nearest-neighbor hopping in two-dimensional
  lattices}, \href{https://doi.org/10.1103/PhysRevB.86.195129}{\emph{Phys. Rev.
  B} {\bfseries 86} (2012) 195129}.

\bibitem{Tadjine:2016:10.1103_PhysRevB.94.075441}
A.~Tadjine, G.~Allan and C.~Delerue, \emph{From lattice hamiltonians to tunable
  band structures by lithographic design},
  \href{https://doi.org/10.1103/PhysRevB.94.075441}{\emph{Phys. Rev. B}
  {\bfseries 94} (2016) 075441}.

\bibitem{Li:2016:10.1039_C6NR03223K}
S.~Li, W.-X.~Qiu and J.-H.~Gao, \emph{Designing artificial two dimensional
  electron lattice on metal surface: a kagome-like lattice as an example},
  \href{https://doi.org/10.1039/C6NR03223K}{\emph{Nanoscale} {\bfseries 8}
  (2016) 12747}.

\bibitem{Wu:2007:10.1103_PhysRevLett.99.070401}
C.~Wu, D.~Bergman, L.~Balents and S.~Das~Sarma, \emph{Flat bands and wigner
  crystallization in the honeycomb optical lattice},
  \href{https://doi.org/10.1103/PhysRevLett.99.070401}{\emph{Phys. Rev. Lett.}
  {\bfseries 99} (2007) 070401}.

\bibitem{Apaja:2010:10.1103_PhysRevA.82.041402}
V.~Apaja, M.~Hyrk\"as and M.~Manninen, \emph{Flat bands, dirac cones, and atom
  dynamics in an optical lattice},
  \href{https://doi.org/10.1103/PhysRevA.82.041402}{\emph{Phys. Rev. A}
  {\bfseries 82} (2010) 041402}.

\bibitem{Shen:2010:10.1103_PhysRevB.81.041410}
R.~Shen, L.B.~Shao, B.~Wang and D.Y.~Xing, \emph{Single dirac cone with a flat
  band touching on line-centered-square optical lattices},
  \href{https://doi.org/10.1103/PhysRevB.81.041410}{\emph{Phys. Rev. B}
  {\bfseries 81} (2010) 041410}.

\bibitem{Guzman-Silva:2014:10.1088_1367-2630_16_6_063061}
D.~Guzm{\'{a}}n-Silva, C.~Mej{\'{\i}}a-Cort{\'{e}}s, M.A.~Bandres,
  M.C.~Rechtsman, S.~Weimann, S.~Nolte et~al., \emph{Experimental observation
  of bulk and edge transport in photonic lieb lattices},
  \href{https://doi.org/10.1088/1367-2630/16/6/063061}{\emph{New Journal of
  Physics} {\bfseries 16} (2014) 063061}.

\bibitem{Mukherjee:2015:10.1103_PhysRevLett.114.245504}
S.~Mukherjee, A.~Spracklen, D.~Choudhury, N.~Goldman, P.~\"Ohberg, E.~Andersson
  et~al., \emph{Observation of a localized flat-band state in a photonic lieb
  lattice}, \href{https://doi.org/10.1103/PhysRevLett.114.245504}{\emph{Phys.
  Rev. Lett.} {\bfseries 114} (2015) 245504}.

\bibitem{Vicencio:2015:10.1103_PhysRevLett.114.245503}
R.A.~Vicencio, C.~Cantillano, L.~Morales-Inostroza, B.~Real,
  C.~Mej\'{\i}a-Cort\'es, S.~Weimann et~al., \emph{Observation of localized
  states in lieb photonic lattices},
  \href{https://doi.org/10.1103/PhysRevLett.114.245503}{\emph{Phys. Rev. Lett.}
  {\bfseries 114} (2015) 245503}.

\bibitem{Taie:2015:10.1126_sciadv.1500854}
S.~Taie, H.~Ozawa, T.~Ichinose, T.~Nishio, S.~Nakajima and Y.~Takahashi,
  \emph{Coherent driving and freezing of bosonic matter wave in an optical lieb
  lattice}, \href{https://doi.org/10.1126/sciadv.1500854}{\emph{Science
  Advances} {\bfseries 1} (2015) e1500854}.

\bibitem{Xia:2016:10.1364_OL.41.001435}
S.~Xia, Y.~Hu, D.~Song, Y.~Zong, L.~Tang and Z.~Chen, \emph{Demonstration of
  flat-band image transmission in optically induced lieb photonic lattices},
  \href{https://doi.org/10.1364/OL.41.001435}{\emph{Opt. Lett.} {\bfseries 41}
  (2016) 1435}.

\bibitem{Diebel:2016:10.1103_PhysRevLett.116.183902}
F.~Diebel, D.~Leykam, S.~Kroesen, C.~Denz and A.S.~Desyatnikov, \emph{Conical
  diffraction and composite lieb bosons in photonic lattices},
  \href{https://doi.org/10.1103/PhysRevLett.116.183902}{\emph{Phys. Rev. Lett.}
  {\bfseries 116} (2016) 183902}.

\bibitem{Slot:2017:10.1038_nphys4105}
M.R.~Slot, T.S.~Gardenier, P.H.~Jacobse, G.C.P.~van Miert, S.N.~Kempkes,
  S.J.M.~Zevenhuizen et~al., \emph{Experimental realization and
  characterization of an electronic lieb lattice},
  \href{https://doi.org/10.1038/nphys4105}{\emph{Nature Physics} {\bfseries 13}
  (2017) 672}.

\bibitem{Drost:2017:10.1038_nphys4080}
R.~Drost, T.~Ojanen, A.~Harju and P.~Liljeroth, \emph{Topological states in
  engineered atomic lattices},
  \href{https://doi.org/10.1038/nphys4080}{\emph{Nature Physics} {\bfseries 13}
  (2017) 668}.

\bibitem{Cui:2020:10.1038_s41467-019-13794-y}
B.~Cui, X.~Zheng, J.~Wang, D.~Liu, S.~Xie and B.~Huang, \emph{Realization of
  lieb lattice in covalent-organic frameworks with tunable topology and
  magnetism}, \href{https://doi.org/10.1038/s41467-019-13794-y}{\emph{Nature
  Communications} {\bfseries 11} (2020) }.

\bibitem{Maldacena:1999:10.1023_A:1026654312961}
J.~Maldacena, \emph{The large-n limit of superconformal field theories and
  supergravity},
  \href{https://doi.org/10.1023/a:1026654312961}{\emph{International Journal of
  Theoretical Physics} {\bfseries 38} (1999) 1113}.

\bibitem{Witten:1998:ATMP.1998.v2.n2.a2}
E.~Witten, \emph{{Anti-de Sitter space and holography}},
  \href{https://doi.org/10.4310/ATMP.1998.v2.n2.a2}{\emph{Adv. Theor. Math.
  Phys.} {\bfseries 2} (1998) 253}
  [\href{https://arxiv.org/abs/hep-th/9802150}{{\ttfamily hep-th/9802150}}].

\bibitem{Gubser:1998:10.1016_S0370-2693(98)00377-3}
S.~Gubser, I.~Klebanov and A.~Polyakov, \emph{Gauge theory correlators from
  non-critical string theory},
  \href{https://doi.org/10.1016/S0370-2693(98)00377-3}{\emph{Physics Letters B}
  {\bfseries 428} (1998) 105}.

\bibitem{Hartnoll:2009:10.1088_0264-9381_26_22_224002}
S.A.~Hartnoll, \emph{Lectures on holographic methods for condensed matter
  physics},
  \href{https://doi.org/10.1088/0264-9381/26/22/224002}{\emph{Classical and
  Quantum Gravity} {\bfseries 26} (2009) 224002}.

\bibitem{Herzog:2009:10.1088_1751-8113_42_34_343001}
C.P.~Herzog, \emph{Lectures on holographic superfluidity and
  superconductivity},
  \href{https://doi.org/10.1088/1751-8113/42/34/343001}{\emph{Journal of
  Physics A: Mathematical and Theoretical} {\bfseries 42} (2009) 343001}.

\bibitem{McGreevy:2010:10.1155_2010_723105}
J.~McGreevy, \emph{Holographic duality with a view toward many-body physics},
  \href{https://doi.org/10.1155/2010/723105}{\emph{Advances in High Energy
  Physics} {\bfseries 2010} (2010) 1}.

\bibitem{Horowitz:2011:10.1007_978-3-642-04864-7_10}
G.T.~Horowitz, \emph{Introduction to holographic superconductors},  in
  \emph{From Gravity to Thermal Gauge Theories: The AdS/CFT Correspondence: The
  AdS/CFT Correspondence}, E.~Papantonopoulos, ed., (Berlin, Heidelberg),
  pp.~313--347, Springer Berlin Heidelberg (2011),
  \href{https://doi.org/10.1007/978-3-642-04864-7_10}{DOI}.

\bibitem{Sachdev:2011:10.1007_978-3-642-04864-7_9}
S.~Sachdev, \emph{Condensed matter and ads/cft},  in \emph{From Gravity to
  Thermal Gauge Theories: The AdS/CFT Correspondence: The AdS/CFT
  Correspondence}, E.~Papantonopoulos, ed., (Berlin, Heidelberg), pp.~273--311,
  Springer Berlin Heidelberg (2011),
  \href{https://doi.org/10.1007/978-3-642-04864-7_9}{DOI}.

\bibitem{Lee:2009:10.1103_PhysRevD.79.086006}
S.-S.~Lee, \emph{Non-fermi liquid from a charged black hole: A critical fermi
  ball}, \href{https://doi.org/10.1103/PhysRevD.79.086006}{\emph{Phys. Rev. D}
  {\bfseries 79} (2009) 086006}.

\bibitem{Liu:2011:10.1103_PhysRevD.83.065029}
H.~Liu, J.~McGreevy and D.~Vegh, \emph{Non-fermi liquids from holography},
  \href{https://doi.org/10.1103/PhysRevD.83.065029}{\emph{Phys. Rev. D}
  {\bfseries 83} (2011) 065029}.

\bibitem{Cubrovic:1993:10.1126_science.1174962}
M.~{\v C}ubrovi{\'c}, J.~Zaanen and K.~Schalm, \emph{String theory, quantum
  phase transitions, and the emergent fermi liquid},
  \href{https://doi.org/10.1126/science.1174962}{\emph{Science} {\bfseries 325}
  (2009) 439}.

\bibitem{Faulkner:2011:10.1098_rsta.2010.0354}
T.~Faulkner, N.~Iqbal, H.~Liu, J.~McGreevy and D.~Vegh, \emph{Holographic
  non-fermi-liquid fixed points},
  \href{https://doi.org/10.1098/rsta.2010.0354}{\emph{Philosophical
  Transactions of the Royal Society A: Mathematical, Physical and Engineering
  Sciences} {\bfseries 369} (2011) 1640}.

\bibitem{Oh:2021:10.1007_JHEP01(2021)053}
E.~Oh, Y.~Seo, T.~Yuk and S.-J.~Sin, \emph{Ginzberg-landau-wilson theory for
  flat band, fermi-arc and surface states of strongly correlated systems},
  \href{https://doi.org/10.1007/jhep01(2021)053}{\emph{Journal of High Energy
  Physics} {\bfseries 2021} (2021) }.

\bibitem{Liu:2018:10.1007_JHEP12(2018)072}
Y.~Liu and Y.-W.~Sun, \emph{Topological nodal line semimetals in holography},
  \href{https://doi.org/10.1007/jhep12(2018)072}{\emph{Journal of High Energy
  Physics} {\bfseries 2018} (2018) }.

\bibitem{Seo:2018:10.1007_JHEP08(2018)077}
Y.~Seo, G.~Song, Y.-H.~Qi and S.-J.~Sin, \emph{Mott transition with holographic
  spectral function},
  \href{https://doi.org/10.1007/jhep08(2018)077}{\emph{Journal of High Energy
  Physics} {\bfseries 2018} (2018) }.

\bibitem{Chakrabarti:2019:10.1007_JHEP07(2019)037}
S.~Chakrabarti, D.~Maity and W.~Wahlang, \emph{Probing the holographic fermi
  arc with scalar field: numerical and analytical study},
  \href{https://doi.org/10.1007/jhep07(2019)037}{\emph{Journal of High Energy
  Physics} {\bfseries 2019} (2019) }.

\bibitem{Oh:2021:10.1007_JHEP11(2021)207}
E.~Oh, T.~Yuk and S.-J.~Sin, \emph{The emergence of strange metal and
  topological liquid near quantum critical point in a solvable model},
  \href{https://doi.org/10.1007/jhep11(2021)207}{\emph{Journal of High Energy
  Physics} {\bfseries 2021} (2021) }.

\bibitem{Laia:2011:10.1007_JHEP11(2011)125}
J.N.~Laia and D.~Tong, \emph{A holographic flat band},
  \href{https://doi.org/10.1007/jhep11(2011)125}{\emph{Journal of High Energy
  Physics} {\bfseries 2011} (2011) }.

\bibitem{Iqbal:2009:10.1002_prop.200900057}
N.~Iqbal and H.~Liu, \emph{Real-time response in ads/cft with application to
  spinors}, \href{https://doi.org/10.1002/prop.200900057}{\emph{Fortschritte
  der Physik} {\bfseries 57} (2009) 367}.

\bibitem{Li:2011:10.1007_JHEP11(2011)018}
W.-J.~Li and H.~Zhang, \emph{Holographic non-relativistic fermionic fixed point
  and bulk dipole coupling},
  \href{https://doi.org/10.1007/jhep11(2011)018}{\emph{Journal of High Energy
  Physics} {\bfseries 2011} (2011) }.

\bibitem{doi:10.1080/23746149.2021.1901606}
J.-W.~Rhim and B.-J.~Yang, \emph{Singular flat bands},
  \href{https://doi.org/10.1080/23746149.2021.1901606}{\emph{Advances in
  Physics: X} {\bfseries 6} (2021) 1901606}
  [\href{https://arxiv.org/abs/https://doi.org/10.1080/23746149.2021.1901606}{{\ttfamily
  https://doi.org/10.1080/23746149.2021.1901606}}].

\bibitem{maimaiti2017compact}
W.~Maimaiti, A.~Andreanov, H.C.~Park, O.~Gendelman and S.~Flach, \emph{Compact
  localized states and flat-band generators in one dimension}, {\emph{Physical
  Review B} {\bfseries 95} (2017) 115135}.

\bibitem{Karch:2006pv}
A.~Karch, E.~Katz, D.T.~Son and M.A.~Stephanov, \emph{{Linear confinement and
  AdS/QCD}}, \href{https://doi.org/10.1103/PhysRevD.74.015005}{\emph{Phys.
  Rev.} {\bfseries D74} (2006) 015005}
  [\href{https://arxiv.org/abs/hep-ph/0602229}{{\ttfamily hep-ph/0602229}}].

\bibitem{Allais:2012ye}
A.~Allais, J.~McGreevy and S.J.~Suh, \emph{{A quantum electron star}},
  \href{https://doi.org/10.1103/PhysRevLett.108.231602}{\emph{Phys. Rev. Lett.}
  {\bfseries 108} (2012) 231602}
  [\href{https://arxiv.org/abs/1202.5308}{{\ttfamily 1202.5308}}].

\bibitem{Hartnoll:2008kx}
S.A.~Hartnoll, C.P.~Herzog and G.T.~Horowitz, \emph{{Holographic
  Superconductors}},
  \href{https://doi.org/10.1088/1126-6708/2008/12/015}{\emph{JHEP} {\bfseries
  12} (2008) 015} [\href{https://arxiv.org/abs/0810.1563}{{\ttfamily
  0810.1563}}].

\end{thebibliography}\endgroup

\end{document}